\documentclass{article}
\usepackage{graphicx}
\usepackage{listings}
\usepackage{fullpage}
\usepackage{fixltx2e}
\usepackage{multirow}
\usepackage{amssymb,amsmath}
\usepackage{mathtools}
\usepackage[hyperfootnotes=false,hidelinks]{hyperref}
\usepackage{url}
\usepackage{subfig}
\usepackage{relsize}
\usepackage{enumitem}
\usepackage{booktabs}
\usepackage{sectsty}
\usepackage{pgffor}
\usepackage{tikz}
\usepackage{cite}
\usepackage{caption}
\usepackage{comment}

\definecolor{myBlueLink}{RGB}{70,70,200}  
\definecolor{myBlue}{RGB}{18,28,43}  
\definecolor{myOrange}{RGB}{26,82,126} 
\definecolor{IllinoisOrange}{RGB}{232,74,39}  
\definecolor{IllinoisBlue}{RGB}{19,41,75}   
\definecolor{mainText}{RGB}{18,28,43}  

\definecolor{ForestGreen}{rgb}{0.0, 0.2, 0.13}
\definecolor{darkpastelgreen}{rgb}{0.01, 0.75, 0.24}
\definecolor{antiquewhite}{rgb}{0.98, 0.92, 0.84}

\def \d {\mathrm{d}}

\def \pp#1#2{\frac{\partial #1}{\partial #2}}

\newcommand{\bu}{\mathbf{u}}

\newcommand{\bbf}{\mathbf{f}}

\newcommand{\DPMsub}{{\text{\textsc{dpm}}}}
\newcommand{\DRLsub}{{\text{\textsc{drl}}}}

\def \Rey {\mathrm{Re}}
\def \Ma {\mathrm{Ma}}
\def \Pr {\mathrm{Pr}}

\newcommand{\vtheta}{{\vec\theta}}

\newcommand{\bQ}{{\mathbf{Q}}}

\definecolor{pdcolor1}{RGB}{23,35,57}    
\definecolor{pdcolor2}{RGB}{184,152,80}  
\definecolor{pdcolor3}{RGB}{60,132,62}  

\definecolor{lc1}{HTML}{CE282A}
\definecolor{lc2}{HTML}{255BCB}
\definecolor{lc3}{HTML}{009647} 
\definecolor{lc4}{HTML}{A21D8E}
\definecolor{lc5}{HTML}{F1AB65}
\definecolor{lc6}{HTML}{89B6D4}
\definecolor{lc7}{HTML}{A0A0A0}

\makeatletter
\renewcommand{\section}{\@startsection
{section}%
{0}%
{0mm}%
{1\baselineskip}%
{0.4\baselineskip}%
{
  \normalfont\Large\bfseries\color{myOrange}}}%
\makeatother

\makeatletter
\renewcommand{\subsection}{\@startsection
{subsection}%
{1}%
{0mm}%
{1\baselineskip}%
{0.4\baselineskip}%
{\normalfont\large\bfseries\color{myOrange}}}%
\makeatother

\makeatletter
\renewcommand{\subsubsection}{\@startsection
{subsubsection}%
{1}%
{0mm}%
{0.1\baselineskip}%
{0.05\baselineskip}%
{\normalfont\normalsize\itshape\centering\color{myOrange}}}%
\makeatother

\newenvironment{noinditemize}
{\begin{itemize}[leftmargin=*,label=\color{myOrange}$\bullet$,itemsep=0in]\raggedright}
{\end{itemize}}

\usepackage{amsmath}
\DeclareMathOperator*{\argmax}{argmax}

\usepackage{algorithm}
\usepackage{algpseudocode}

\newcommand\figref{Figure~\ref}

\usepackage{xargs}                      
\usepackage[colorinlistoftodos,prependcaption,textsize=tiny]{todonotes}
\newcommandx{\unsure}[2][1=]{\todo[linecolor=red,backgroundcolor=red!25,bordercolor=red,#1]{#2}}
\newcommandx{\clarify}[2][1=]{\todo[linecolor=red,backgroundcolor=red!25,bordercolor=red,#1]{#2}}
\newcommandx{\change}[2][1=]{\todo[linecolor=red,backgroundcolor=red!25,bordercolor=red,#1]{#2}}
\newcommandx{\info}[2][1=]{\todo[linecolor=lc5,backgroundcolor=lc5!25,bordercolor=lc5,#1]{#2}}
\newcommandx{\improvement}[2][1=]{\todo[linecolor=lc3,backgroundcolor=lc3!25,bordercolor=lc3,#1]{#2}}

\usepackage[export]{adjustbox}
\usepackage[algo2e]{algorithm2e}

\title{Adjoint-based machine learning for active flow control}

\author{Xuemin Liu\footnote{xliu24@nd.edu}\phantom{,} and Jonathan F.~MacArt\footnote{jmacart@nd.edu} \\
  {\normalsize Department of Aerospace and Mechanical Engineering, University of Notre Dame}}

\begin{document}

\maketitle





\thispagestyle{plain}
\begin{abstract}

We develop neural-network active flow controllers using a deep learning PDE augmentation method (DPM). The end-to-end sensitivities for optimization are computed using adjoints of the governing equations without restriction on the terms that may appear in the objective function.
In one-dimensional Burgers' examples with analytic (manufactured) control functions, DPM-based control is comparably effective to standard supervised learning for in-sample solutions and more effective for out-of-sample solutions, i.e., with different analytic control functions. The influence of the optimization time interval and neutral-network width are analyzed, the results of which influence algorithm design and hyperparameter choice, balancing control efficacy with computational cost. We subsequently develop adjoint-based controllers for two flow scenarios.
First, we compare the drag-reduction performance and optimization cost of adjoint-based controllers and deep reinforcement learning (DRL)-based controllers for two-dimensional, incompressible, confined flow over a cylinder at $Re=100$, with control achieved by synthetic body forces along the cylinder boundary. The required model complexity for the DRL-based controller is 4,229 times that required for the DPM-based controller.
In these tests, the DPM-based controller is 4.85 times more effective and 63.2 times less computationally intensive to train than the DRL-based controller.
Second, we test DPM-based control for compressible, unconfined flow over a cylinder and extrapolate the controller to out-of-sample Reynolds numbers. We also train a simplified, steady, offline controller based on the DPM control law. Both online (DPM) and offline (steady) controllers stabilize the vortex shedding with a 99\% drag reduction, demonstrating the robustness of the learning approach. For out-of-sample flows ($Re=\{50,200,300,400\}$), both the online and offline controllers successfully reduce drag and stabilize vortex shedding, indicating that the DPM-based approach results in a stable model. A key attractive feature is the flexibility of adjoint-based optimization, which permits optimization over arbitrarily defined control laws without the need to match \emph{a priori} known functions.


\end{abstract}
\vspace{1em}

\section{Introduction}

Altering the natural dynamics of flows via control is desirable for engineering applications\cite{ashill2001flow}.  Examples include drag reduction to reduce aircraft fuel consumption \cite{sudin2014review, wang2019investigation} and structural damage prevention by suppressing flow-induced oscillations \cite{CATTAFESTA2008479}.
Flow control can be broadly classified into passive and active methods.
Passive flow control (PFC) requires no external energy, relying on inherent characteristics of the flow system  or the use of passive devices, while active flow control (AFC) requires energy inputs.
Typical PFC techniques include geometric modifications, for example, trailing-edge flaps on airfoils to reduce nose-down pitching moments \cite{gerontakos2006dynamic}, or surface roughness designed to reduce skin friction drag \cite{bliamis2022numerical}. Aerodynamic shape optimization is closely related, being typically fixed in operation, for example, optimizing over computational fluid dynamics (CFD) calculations to increase the lift-to-drag ratio \cite{acarer2020peak, SRINATH20101994}. 
Though passive flow control techniques are cost-effective, their control efficacy is limited and is sensitive to fouling and damage, which may not be suitable for all applications. 
Conversely, active flow control can offer greater flexibility, though with the requirement of actuation energy, and has been a fixture of aerodynamics since Prandtl's pioneering work on boundary-layer separation delay via oscillatory blowing and suction~\cite{prandtl1904flussigkeitsbewegung}.

AFC methods can be categorized into predetermined (open-loop) and interactive (closed-loop) control. 
Open-loop control involves energy expenditure without necessarily measuring the flow field. 
Closed-loop control instead modulates actuators using sensor measurements; recent examples include blowing/suction jets to reduce drag over 2D cylinders \cite{rabault2019artificial, tang2020robust} and  hot-wire sensors and loudspeaker actuators to stabilize the wake instability \cite{roussopoulos1993feedback}. Choi et al.~\cite{choi1993feedback} proposed an effective framework for turbulent flows, and other critical theoretical developments can be found in the review by Brunton and Noack~\cite{brunton2015closed}.
Combinations of PFC and AFC have also been presented, such as using a synthetic jet actuator and porous coatings to adjust drag and lift acting on a cylinder \cite{act11070201}. In general, active flow control provides more advanced and effective flow control  than passive techniques.

Nevertheless, developing efficient AFC strategies remains a challenge~\cite{SCOTTCOLLIS2004237}. 
Machine learning (ML) with artificial neural networks (ANNs) has received significant attention, for it is well suited to optimization and control problems involving black-box or multimodal cost functions~\cite{brunton2020machine}.
One approach is to construct surrogate models for the flow dynamics using data-driven ML methods; examples include predicting drag and lift by training convolutional neural networks from CFD simulations \cite{portal2022cnn}, learning a multi-fidelity approximator for flow simulations \cite{liao2021multi}, and reduced-order modeling of flow dynamics \cite{mohan2018deep,otto2019linearly}. 

Another popular branch of ML for AFC utilizes deep reinforcement learning (DRL), which models an agent interacting with its environment so as to maximize the cost function based on a Markov decision process.
DRL is widely used for complex decision-making problems (originally associated with games) and has been applied to an increasing number of physical systems. In the past few years, DRL has been applied to laminar drag reduction using synthetic blowing and suction \cite{rabault2019artificial}, shape optimization of airfoils~\cite{viquerat2021direct}, and turbulent drag reduction in channel flows via  blowing/suction controlled by velocity measurements~\cite{sonoda2023reinforcement} and the wall shear stress~\cite{lee2023turbulence}.
Other algorithms have been combined with DRL to help discover better control policies; for example,
sparse proximal policy optimization with covariance matrix adaptation (S-PPO-CMA) has been used to optimize sensor layouts for drag reduction \cite{paris2021robust} and to augment with RL to suppress vortex shedding~\cite{li2022reinforcement}.
Pino et al.~\cite{pino2023comparative} compared the AFC performance of multiple ML methods including generic programming (GP), deep deterministic policy gradient (DDPG, a DRL variant),
Lipschitz global optimization (LIPO), and Bayesian optimization (BO).
More recent applications can be found in the review by Vignon et al.~\cite{vignon2023recent}.
Though DRL is attractive for AFC applications only requiring a proper-defined environment without the programming optimization algorithms (provided by many open-sourced libraries), there are still many challenges because of the sample inefficiency, difficulty of designing reward functions, and lack of stability and convergence theories  \cite{li2017deep}. 



Gradient-based optimization can also be used to develop active flow controllers. It requires gradients of a cost function $L(\bu,\vtheta)$ to be computed with respect to control parameters $\vtheta$,
\begin{equation} \label{eq:gradvec}
      \nabla_{\theta} {L(\bold{u}, \vec{\theta})} = \pp{L}{\bold{u}} \pp{\bold{u}}{\vec{\theta}} + \pp{L}{\vec{\theta}},
\end{equation}
where $\bu$ are the flow variables that implicitly depend on $\vtheta$. For systems governed by partial differential equations (PDEs), $\bu$ is often high-dimensional, making $\partial{\bold{u}} / \partial{\vec{\theta}}$ computationally prohibitive to compute directly, especially when $\vec{\theta}$ is also high-dimensional.
Adjoint-based methods, which we pursue, enable efficient computation of $\nabla_{\theta} {L}$ at expense independent of the dimension of $\vtheta$~\cite{carnarius2013optimal}. 
Adjoint methods can be broadly classified into continuous and discrete approaches: in the continuous approach, the optimization problem is first stated in continuous form, then discretized, while the discrete approach first discretizes the forward problem, then poses the optimization problem in discrete form.
Adjoint-based optimization originates from dynamic programming methods in optimal control theory \cite{bellman1956dynamic}. Optimization and control are closely related, and their boundaries are becoming less distinct with increasingly capable computers \cite{tsiotras2017toward}.
Applications of adjoint-based methods in aerodynamics and flow control include shape optimization of aircraft to increase the lift-to-drag ratio \cite{jameson1988aerodynamic, reuther1999constrained, SRINATH20101994}, topology optimization for unsteady incompressible fluid flows \cite{gorodetsky2018gradient},
rotating-cylinder drag reduction \cite{carnarius2010adjoint}, and multi-mode Rayleigh–Taylor (RT) instability suppression~\cite{kord2019Optimal}.

  
A deep learning PDE augmentation method (DPM) was recently proposed to train ML models using adjoint-based, PDE-constrained optimization. While successful for large-eddy simulation (LES) subgrid-scale modeling \cite{Sirignano2020DPM,MacArt2021Embedded}, it has not yet been applied to flow-control tasks. 
A key feature of the DPM is its use of adjoints to provide the end-to-end sensitivities needed for ML model optimization; thus, the models it trains are constrained by the system's dynamics, at least as accurately as the dynamics are modeled by the forward PDEs (unlike DRL, which approximates the system's dynamics using additional ML models).  The convergence of DPM to a global minimum during optimization has been proved for linear elliptic PDEs with neural network terms \cite{sirignano2023pde}, and further study of its application to linear and nonlinear parabolic and hyperbolic PDEs is ongoing.
The present work develops a DPM approach to AFC problems, in which the parameters of a flow controller are optimized by solving the flow PDEs, subject to a user-defined objective function.



This paper is organized as follows.
Section~\ref{sec:method} introduces the DPM and DRL optimization methods and verifies the convergence of DPM-based flow controllers for the 1D viscous Burgers' equation.
Section~\ref{sec:DPMvsDRL} evaluates and compares the control performance and training cost of DPM- and DRL-based active flow controllers for 2D incompressible flow over a confined cylinder.
Section~\ref{sec:PyflowCLDPM} analyzes the performance of DPM-trained controllers for 2D compressible flow over an unconfined cylinder, assesses the ability of the learned controller to extrapolate to higher Reynolds numbers, and tests the performance of a simplified controller inferred from the DPM-learned controller. A summary and discussion are given in Section~\ref{sec:Conclusion}

\section{Optimization-based flow control} \label{sec:method}

Sections \ref{subsec:DRL} and \ref{sec:DPM} introduce the DRL and DPM algorithms and their applications to flow control. Section~\ref{sec:burgers} verifies and analyzes the convergence of the DPM-based control strategy for a 1D viscous Burgers' equation example.

\subsection{Deep reinforcement learning} \label{subsec:DRL}
\begin{figure}[H]
    \centering
    \includegraphics[width=0.4\linewidth]{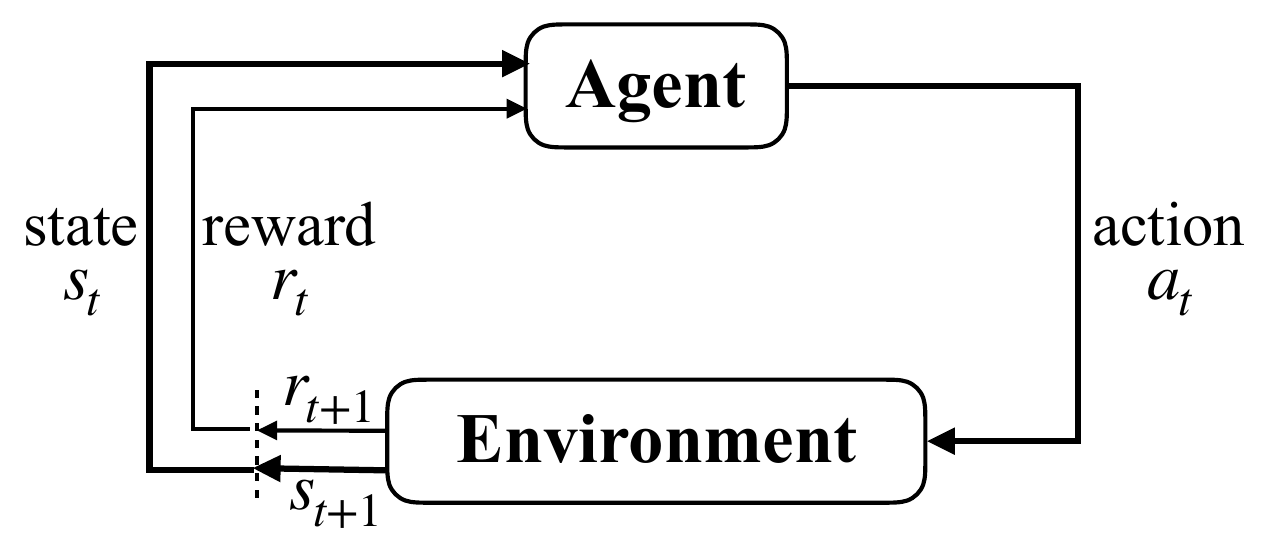}
    \caption{DRL: interaction of the agent and environment in a Markov decision process.}
    \label{fig:MDP}
\end{figure}

DRL constructs a model for an agent interacting with an environment so as to maximize an objective\cite{SuttonBarto2018rl}, as shown in Figure~\ref{fig:MDP}.
The agent is a decision-maker that maps observed states and rewards (feedback) to actions:  $(s_t, r_t) \mapsto a_t$. The environment is a dynamical system that advances to the next state $s_{t+1}$ with new feedback $r_{t+1}$.
This is mathematically based on a Markov decision process (MDP), in which the transitions between states depend only on the current state and action (not on any previous states or actions).
In general, its goal is to maximize the cumulative reward $R(\mathcal{T}) = \sum_{t=0}^{T}\gamma^tr(s_t,a_t)$ along a trajectory  $\mathcal{T} = (s_0,a_0,r_0; s_1,a_1,r_1;\ldots;  s_T,a_T,r_T)$ 
with a discount rate $\gamma \in [0,1]$ that smooths the impact of temporally distant rewards \cite{SuttonBarto2018rl}.
DRL algorithms are classified into policy-based and  value-based methods.
In policy-based methods, the agent approximates a policy $\pi(a|s)$ that explicitly maps states to actions, where $\pi(a|s) = \mathrm{Pr}(a| s)$ is the  probability of action $a$ given the observed state $s$.
In contrast, value-based methods obtain the action at time $t$ implicitly by $a_t = \argmax_a V^{\pi}(s_{t+1})$, where $V^{\pi}(s) =  \mathbb{E}_{\mathcal{T} \sim \pi}[R(\mathcal{T})|s]$ is a state value function, 
and $\mathbb{E}_{\mathcal{T} \sim \pi}[\cdots]$ refers to the ensemble mean over several stochastic trajectories $\mathcal{T}$, along which actions are randomly sampled from the policy. Actor-critic algorithms are a combination of these, where policy and value functions are learned simultaneously~\cite{konda1999actor}. 

For comparisons to adjoint-based learning, we employ an actor-critic proximal policy optimization (PPO) algorithm to obtain an optimal policy $\pi^{*}(a|s)$; this approach is known to have higher data efficiency, robustness, and simplicity than other DRL methods~\cite{schulman2017proximal}. 
The environment is a flow simulation (PDE solution), and the discovery of the PPO agent is done using the open-source \textit{Tensorforce} library \cite{tensorforce2017}.
Pseudocode for the PPO algorithm  is given in Algorithm \ref{alg:PPO}  to maximize an objective function $\mathcal{L}$, 
\begin{equation} \label{eq:ppoL}
    \mathcal{L}(\theta) = \min
            \left(
            \frac{\pi_{\theta}(a_t|s_t)}{\pi_{\theta_{k}}(a_t|s_t)} A^{\pi_{\theta_k}}(s_t,a_t), \ \ 
            \mathrm{clip}\Big( \frac{\pi_{\theta}(a_t|s_t)}{\pi_{\theta_{k}}(a_t|s_t)}, 1 - \varepsilon, 1 + \varepsilon \Big)
            A^{\pi_{\theta_k}}(s_t,a_t) \right ),
\end{equation}
where $\pi_{\theta}$ is a stochastic policy between  steps $k$ and $k+1$, $A^\pi$ is the estimated advantage value corresponding to the  policy, and $\varepsilon$ is a (small) hyperparameter that determines the maximum permitted distance of the new policy  from the old policy.
During testing, the action is ensured to be deterministic by directly using the mean as the proposed action (i.e., disregarding any randomness).
More details of this algorithm can be found in \cite[Section 3]{schulman2017proximal} and \cite[Appendix C]{rabault2019artificial}.
\begin{algorithm}[H]
  \SetAlgoLined
   Initialize actor network $\theta_0$ and critic network $\phi_0$\\
  \For{$k \gets 1$ \KwTo $K$}{
  \For{$n \gets 1$ \KwTo $N$}{
    Run policy $\pi_{\theta_k}$ in environment for $T$ timesteps\\
    Compute cumulative reward $R_t=\sum_{i=t}^{T} \gamma^{n-i}r_i$ \\
    Estimate advantage $A_t^{\pi_{\theta_k}} (s_t,a_t)= R_t-V_{\phi_{k}}(s_t) $ by evaluating the value function $V_{\phi_k}(s_t)$ 
    }
    Optimize the objective function $\mathcal{L}$ wrt $\theta$, then $\theta_{k+1} \gets \theta$  \\
    Optimize value function by minimizing the regression loss
    ${\frac{1}{NT} \sum_{n=0}^{N} \sum_{t=0}^{T}(V_{\phi}(s_t) -  R_t)^2  }$, then $\phi_{k+1} \gets \phi$
    }
  \caption{Actor-critic style PPO} \label{alg:PPO}
\end{algorithm}

While DRL is a popular method for decision-making problems, its drawbacks include sample inefficiency, difficulty of designing a successful reward function, and lack of proofs of stability and convergence. We return to these challenges in Section~\ref{sec:DPMvsDRL}, when we compare DRL to DPM for flow control.

\subsection{Deep learning-based PDE augmentation} \label{sec:DPM}

Rather than attempt to approximate the system dynamics with a neural network, DPM optimizes directly over the governing PDEs.
The discrete solution of a system of PDEs with embedded neural-network terms, $u(t,\theta) \in \mathbb{R}^{n_d}$,  is an implicit function of the neural-network parameters $\theta\in\mathbb{R}^{n_\theta}$, where $n_d=d\times n_\mathrm{eq}$ for $d$ discrete mesh nodes and $n_\mathrm{eq}$ dependent variables. For typical systems encountered in engineering and applied science, $n_d=O(10^5)$ to $O(10^9)$, and $n_\theta$ can be $O(10^5)$ or larger for modern ML models. Optimizing $\theta$ using standard gradient-descent methods would require calculating $\nabla_\theta u \in\mathbb{R}^{n_d\times n_\theta}$, which would be prohibitively expensive for typical physical systems. DPM instead calculates the gradients needed for optimization by solving adjoint variables $\hat u(t)\in\mathbb{R}^{n_d}$ with comparable cost to the solution of the forward PDE system.

We wish to minimize a time-averaged objective function of the PDE solution
 \begin{equation}\label{eq:integObj}
        \Bar{J} = \int_0^\tau J(u,t, \theta) dt ,
    \end{equation}  
where $J(u,t,\theta)$ is an instantaneous scalar objective function, and
$\tau$ is the time interval over which optimization is performed.
Given initial conditions $g(u(0))=0$ satisfying the PDE and an instantaneous PDE residual $R(u,\dot u, t, \theta)$, where $\Dot{u} = \partial u/\partial t$, the optimization problem is
    \begin{equation}\label{eq:obj}
       \min \Bar{J} \mathrm{\ over\ } (u, \theta) \mathrm{\ subject\ to\  } R(u, \Dot{u}, t, \theta) = 0 \mathrm{\ and\ } g(u(0)) = 0.
     \end{equation}
The ANN parameters are updated from time level $n$ to time level $n+1$ using a gradient descent step,
    \begin{equation} \label{eq:thetaupdate}
        \theta_{n+1} = \theta_n - \alpha_n \nabla_{\theta_n} \Bar{J},
    \end{equation}
where $\alpha_n$ is the learning rate at step $n$. DPM computes $\nabla_{\theta} \Bar{J}$ using the adjoint variables $\hat u(t)$,
    \begin{equation} \label{eq:gradL}
            \nabla_{\theta} \Bar{J} = 
           \int_{0}^{\tau} \nabla_{\theta} J  dt =
            \int_{0}^{\tau} \left( \Hat{u}^\top \pp{R}{\theta} + \pp{J}{\theta} \right) dt,
    \end{equation}
    where $\hat u(t)$ satisfy the linear ordinary differential equation (ODE)
    \begin{equation} \label{eq:uhat}
         {\color{black}\frac{d\Hat{u}^\top}{dt}}  
         =   \Hat{u}^\top \pp{R}{u} 
         + \pp{J}{u},
    \end{equation}
which is solved backward in time from $t=\tau$ to $t=0$ with initial condition $\hat u(\tau)=0$. In \eqref{eq:uhat}, the Jacobian $\partial R/\partial u$ can be challenging to obtain in general; we compute it using automatic differentiation (AD) over the the forward solution using the  \textit{PyTorch} library \cite{NEURIPS2019_9015}.
AD works by sequentially applying the chain rule of differentiation to evaluate the gradient of a function with respect to its inputs.

\begin{figure}[H]
    \centering
    \includegraphics[width=0.45\linewidth]{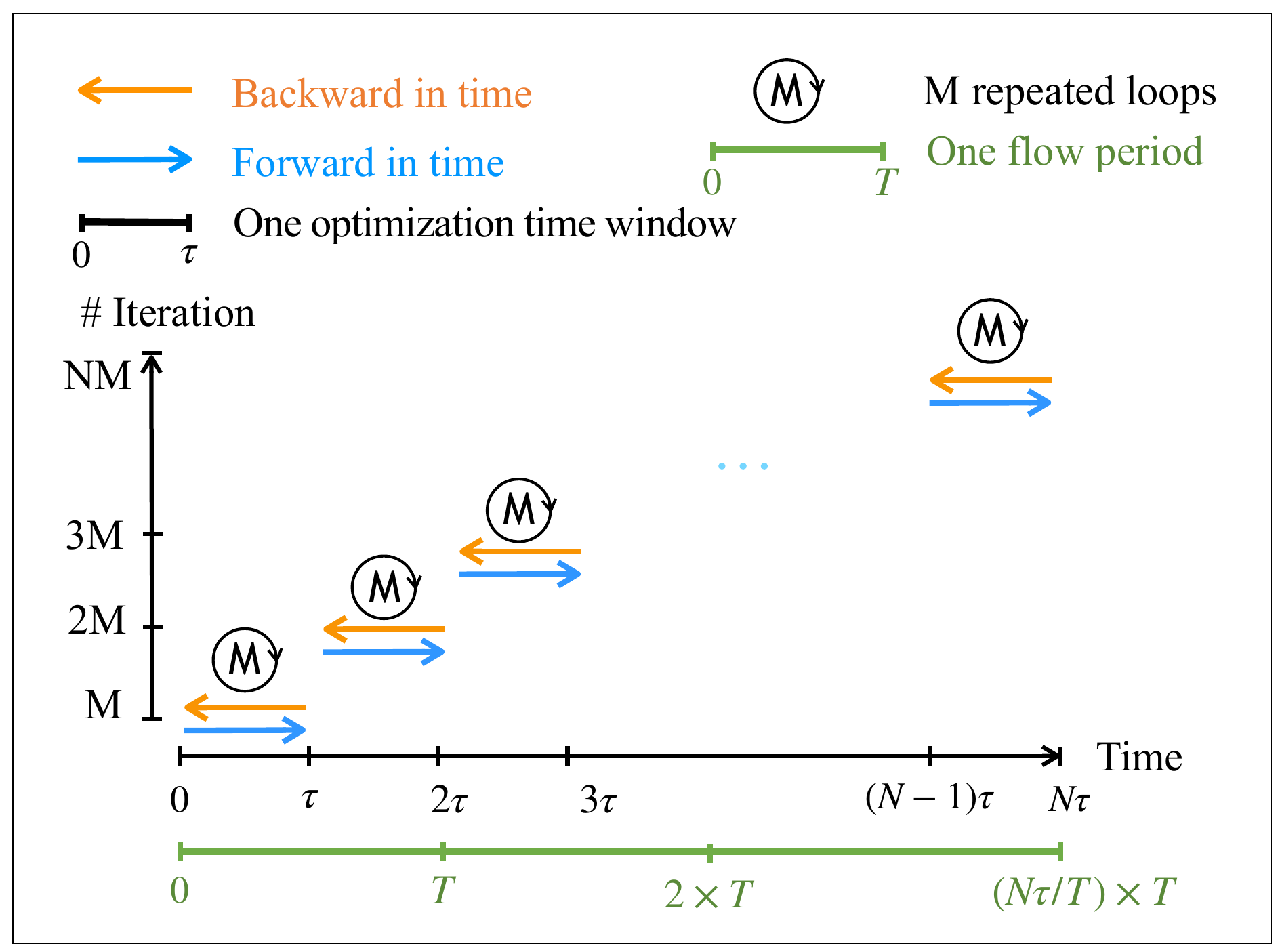}
    \caption{Illustration of the DPM optimization steps in Algorithm \ref{alg:DPM}, with $\tau=T/2$ as an example.}
    \label{fig:NMtauT}
\end{figure}

Numerical stability often requires a system's characteristic time $T$ to be discretized over many time steps. Solving \eqref{eq:uhat} requires checkpointed solutions $u(t,\theta)$ over $t\in[0,\tau]$ at intermediate time steps, with available computer memory often dictating the maximum ratio $\tau/T$. This is an important consideration for flows, for the maximum $\tau/T$ can be significantly less than unity. Figure~\ref{fig:NMtauT} illustrates the decomposition of multiple characteristic times $T$ into $N$ optimization windows of size $\tau = T/2$ (for example), each of which is optimized $M$ times for a total of $NM$ optimization iterations. Pseudocode for the DPM algorithm is given in Algorithm \ref{alg:DPM}.




\begin{algorithm}[H]
  \SetAlgoLined
   Initialize ANN with initial parameters $\theta_0$.
   Set the time interval for optimization $\tau$ and total number of optimization steps $NM$\\
   \For{$n \gets 0$ \KwTo $N$}{
      \For{$m\gets 0$ \KwTo $M$}{
          \For{$i \gets n\tau$ \KwTo $(n+1)\tau$}{
            Compute the flow variables $u_i$ by simulating the PDE $R(u, \Dot{u}, \theta,t) = 0$ forward in time
            }
          Initialize adjoint variables $\hat{u}_{(n+1)\tau}=0$\\
          \For{$j \gets (n+1)\tau$ \KwTo $n\tau$}{
            Compute the adjoint variables $\hat{u}_j$ backward in time by solving Eq.~\eqref{eq:uhat}
            }
        Compute gradients $\nabla_{\theta} \Bar{J}$ via Eq.~\eqref{eq:gradL}\\
        Optimize the controller $\theta \gets \theta^{'}$  via Eq.~\eqref{eq:thetaupdate}
        }
    }
  \caption{DPM} \label{alg:DPM}
\end{algorithm}

\subsection{Example: Burgers’ equation control} \label{sec:burgers}

We now demonstrate DPM-based control of the viscous Burgers' equation with the control objective provided by an analytic solution. The Burgers' equation has a similar nonlinearity to the Navier--Stokes equation, hence it is widely used as a numerical test  \cite{whitham2011linear}. We prescribe an analytic solution and obtain analytic control terms using the method of manufactured solutions (MMS), obtain approximate control terms using DPM and \emph{a priori} ML, and compare the learned controllers to the analytic controller.

\subsubsection{Control framework} \label{sec:control_framework}

Consider the one-dimensional viscous Burgers’ equation with a source term,
\begin{align}
  \pp{u}{t} + u\pp{u}{x} = \nu \pp{^2 u}{x^2} + S(\bu,\theta),  \label{eq:burgers}
\end{align}
with domain $x\in[0,L]$, $L=1$, solution $u(x,t)$, initial condition $u_0(x)$, and periodic boundary conditions. The source term has inputs $\bu$ and, in the case of modeled source terms, tunable parameters $\theta$. A constant viscosity $\nu=0.002$ is used. MMS prescribes a target solution $u^e(x,t)$ for which an analytic source term $S^e(x,t)$ may be obtained. Modeling the source term using a neural network enables direct comparison of ML-based control methods to the exact control term.

We aim to learn a class of controllers $S(\bu,\theta)$ that modify the dynamics of \eqref{eq:burgers} to match those of a first-order linear advection equation,
\begin{align}
  \pp{u}{t} + a\pp{u}{x} = 0, \label{eq:adv}
\end{align}
with matching domain, initial condition, and boundary conditions as \eqref{eq:burgers} and constant advective velocity $a$, giving a characteristic time scale $T=L/a$.
The analytic source term for a prescribed $u^e$ is therefore
\begin{equation}
  S^e = (u^e-a) \pp{u^e}{x} - \nu \pp{^2 u^e}{x^2}. \label{eq:MMSsource}
\end{equation}
We represent the modeled source term $S(\bu,\theta)$ using a simple, four-layer, fully connected neural network with 200 hidden units per layer. The neural network has the same architecture as one used previously for LES subgrid modeling \cite{Sirignano2020DPM}.
At a mesh node $x_i$, the neural network inputs are the spatially local solution $u_i=u(x_i)$ and its nearest neighbors $\bu=(u_{i-3},u_{i-2},u_{i-1}, u_i,u_{i+1},u_{i+2},u_{i+3})$, and its output is the source term at the mesh node $x_i$.

We consider two approaches to optimize the controller parameters $\theta$. Both minimize a time-integrated loss $\Bar{J} = \int_0^\tau  J dt$ but differ in the choice of instantaneous loss $J$ and hence how the optimization is performed.
\begin{enumerate}
\item \textbf{\emph{A priori} ML} minimizes $J(\theta) = \frac{1}{2} (S(\bu^e,\theta) - S^e)^\top(S(\bu^e,\theta) - S^e)$, where the modeled $S(\bu^e,\theta)$ is evaluated using the \emph{a priori}-known exact solution $\bu^e$. This can be performed offline, without solving \eqref{eq:burgers}, hence is computationally inexpensive.

\item \textbf{DPM (adjoint-based ML)} minimizes $J(u(\theta)) = \frac{1}{2} (u(\theta) - u^e)^\top(u(\theta) - u^e)$ by optimizing over \eqref{eq:burgers}, where we explicitly denote the dependence of the computed solutions $u$ on $\theta$. This is done using the adjoint method described in Section~\ref{sec:DPM} with residual $R = \pp{u}{t} + u\pp{u}{x} - \nu \pp{^2 u}{x^2} - S(\bu,\theta)$. 
\end{enumerate}

For model training, we consider a Gaussian initial profile $u_0(x) = \exp\left[-(x-x_0)^2/(2\sigma^2)\right]$, where $\sigma=0.1$ and  $x_0=L/2$.
For numerical tests, the computational domain is discretized using a uniform mesh of 256 grid points, and all derivatives are calculated using second-order central differences. Time is advanced using the  forward Euler method with step size $\Delta t=2\times10^{-4}$, which corresponds to a CFL number of approximately 0.05.
The learning rate is initialized as $\alpha_0=10^{-3}$ and decays as $\alpha_{n}=0.997^n\alpha_0$ for \emph{a priori} and  $\alpha_{n}=0.997^{\lfloor n/5 \rfloor}\alpha_0$ for DPM.

\subsubsection{\emph{A priori} ML versus DPM control performance} \label{sec:dpm_vs_apriori}

We first compare the in- and out-of-sample performance of \emph{a priori} ML and DPM control, where the DPM-trained controllers use a fixed training window $\tau=100\Delta t$. The influence of the training window $\tau$ is assessed in Section~\ref{subsubsec:tau/T} for in-sample cases.

\begin{figure}
\centering
\includegraphics[width=1.0\linewidth]{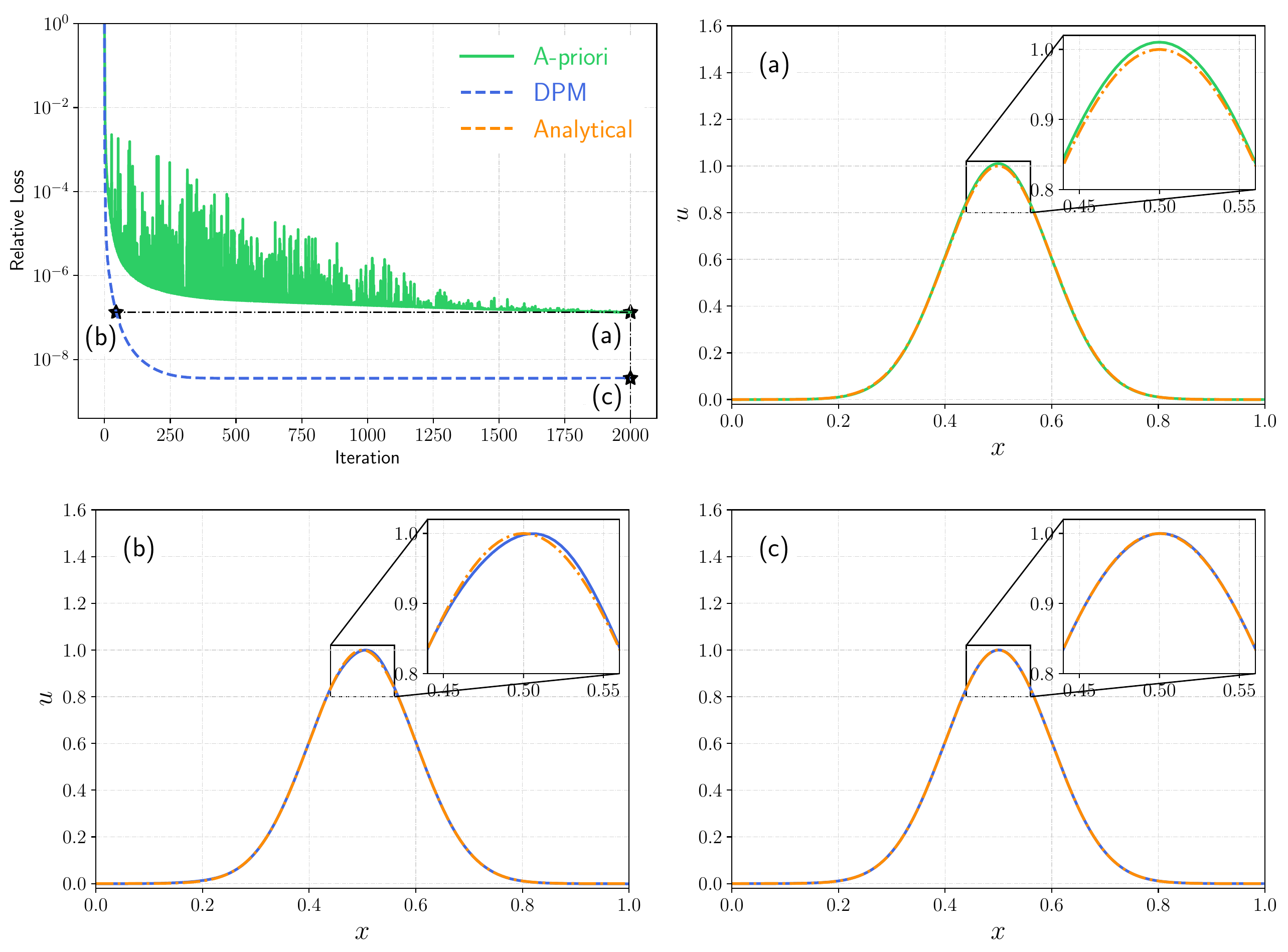}
\caption{Top left: Relative training loss for \emph{a priori} and DPM-trained models $\epsilon_\mathrm{rel}=J_n/J_0$, where $J_0$ is the norm of targets. Remaining quadrants: at $t/T=1$, (a) Solution using best-case \emph{a priori} ML model trained for 2000 iterations; (b) Solution using DPM model with matching training relative loss (43 iterations); (c) Solution using DPM model trained for 2000 iterations.}
\label{fig:loss_cmp}
\end{figure}



The training dataset for \emph{a priori} ML models comprises source-term snapshots $S^e(x,i\Delta t)$, $i=1,\dotsc,5000$, which covers $t/T\in[0,1]$ for $\Delta t=2\times10^{-4}$.  This process requires 71 minutes on one NVIDIA RTX-6000 GPU for 2000 training iterations. 
Optimization of the DPM models targets $u^e(x,i\Delta t)$, $i=1,\dotsc,5000$ using the same learning rate and exponential scheduler. For the same number of optimization iterations, DPM training requires 964 minutes ($13\times$ the times computational cost of \emph{a priori} training). This increase is understandable, for DPM training includes the cost of solving the governing equations.

Figure~\ref{fig:loss_cmp} compares the relative training loss of \emph{a priori} and DPM-trained models, $\epsilon_\mathrm{rel}=J_n/J_0$, where $J_n$ is the per-iteration loss, and $J_0$ is the norm of targets $J_0(\theta) = \frac{1}{2} (S^e)^\top(S^e)$ for DRL and $J_0(u(\theta)) = \frac{1}{2} (u^e)^\top(u^e)$ for DPM. Since \emph{a priori} ML and DPM use different loss functions (Section~\ref{sec:control_framework}), the relative loss is the appropriate metric to compare the relative convergence of trained models. The \emph{a priori}-trained model converges to $\epsilon_\mathrm{rel}=1.34\times10^{-7}$ over 2000 training iterations. The DPM-trained model achieves $\epsilon_\mathrm{rel}=O(10^{-9})$ convergence in as few as 250 iterations and converges to $\epsilon_\mathrm{rel}=2.71\times10^{-9}$ over 2000 iterations. To obtain $\epsilon_\mathrm{rel}=O(10^{-7})$, matching the \emph{a priori} model's convergence after 2000 iterations, the DPM model requires only 43 training iterations. The wall-time for this was 20.7 minutes---approximately one-third the training time of the \emph{a priori} model.

Three trained models, indicated in the top-left quadrant of Figure~\ref{fig:loss_cmp}, are chosen for testing: (a) an \emph{a priori} model trained for 2000 iterations, (b) a DPM model trained for 43 iterations, matching the minimum \emph{a priori} relative training error, and (c) a DPM model trained for 2000 iterations. Instantaneous snapshots of the controlled solutions for each of these three models at $t/T=1$ are shown in the remaining quadrants of Figure~\ref{fig:loss_cmp}. At $t/T=1$, models (a) and (b) have $O(10^{-5})$ \emph{a posteriori} testing error, while  the error for model (c) is $O(10^{-7})$ and is visually indistinguishable from the analytical target solution.

\begin{figure}
\centering
\includegraphics[width=0.48\linewidth]{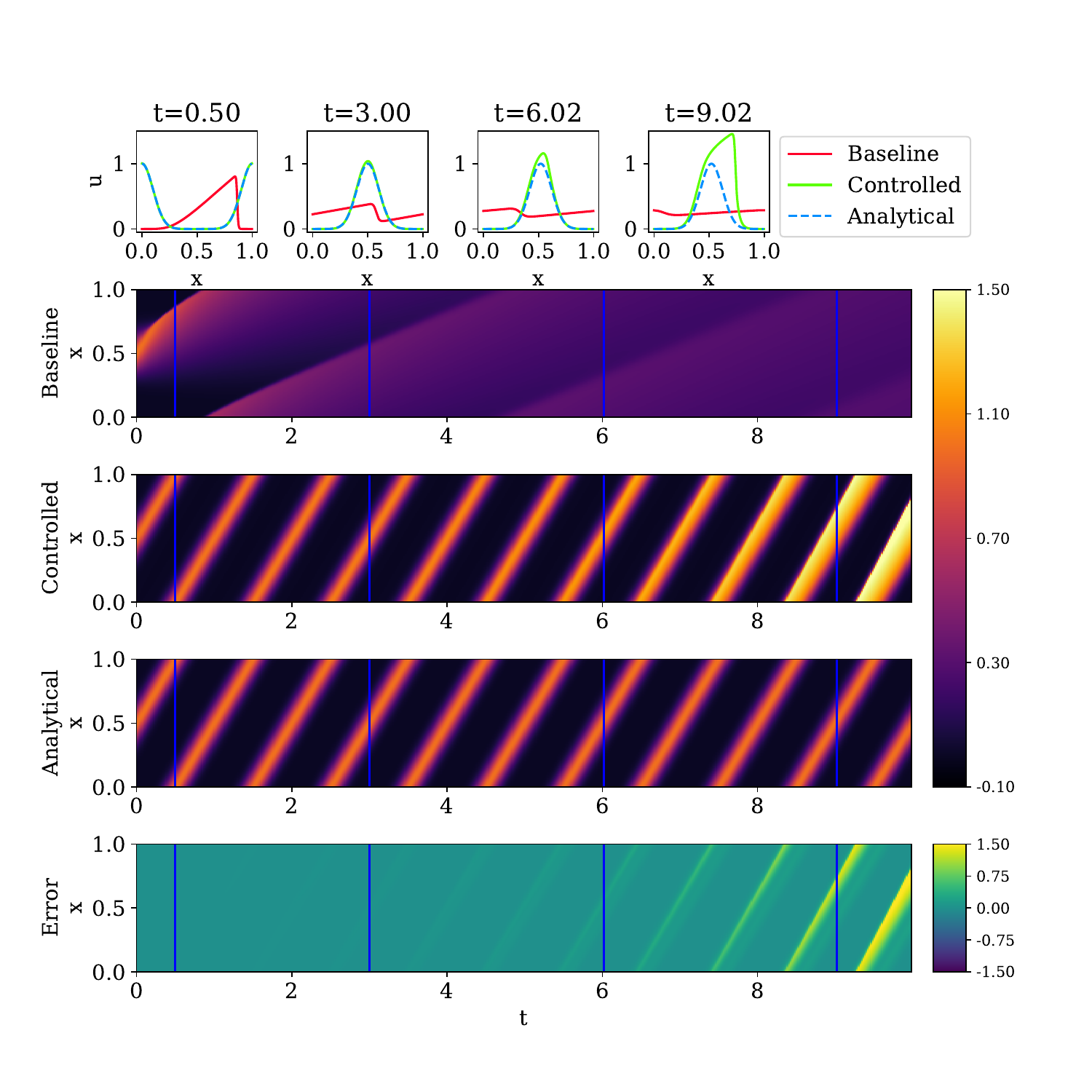}
\includegraphics[width=0.48\linewidth]{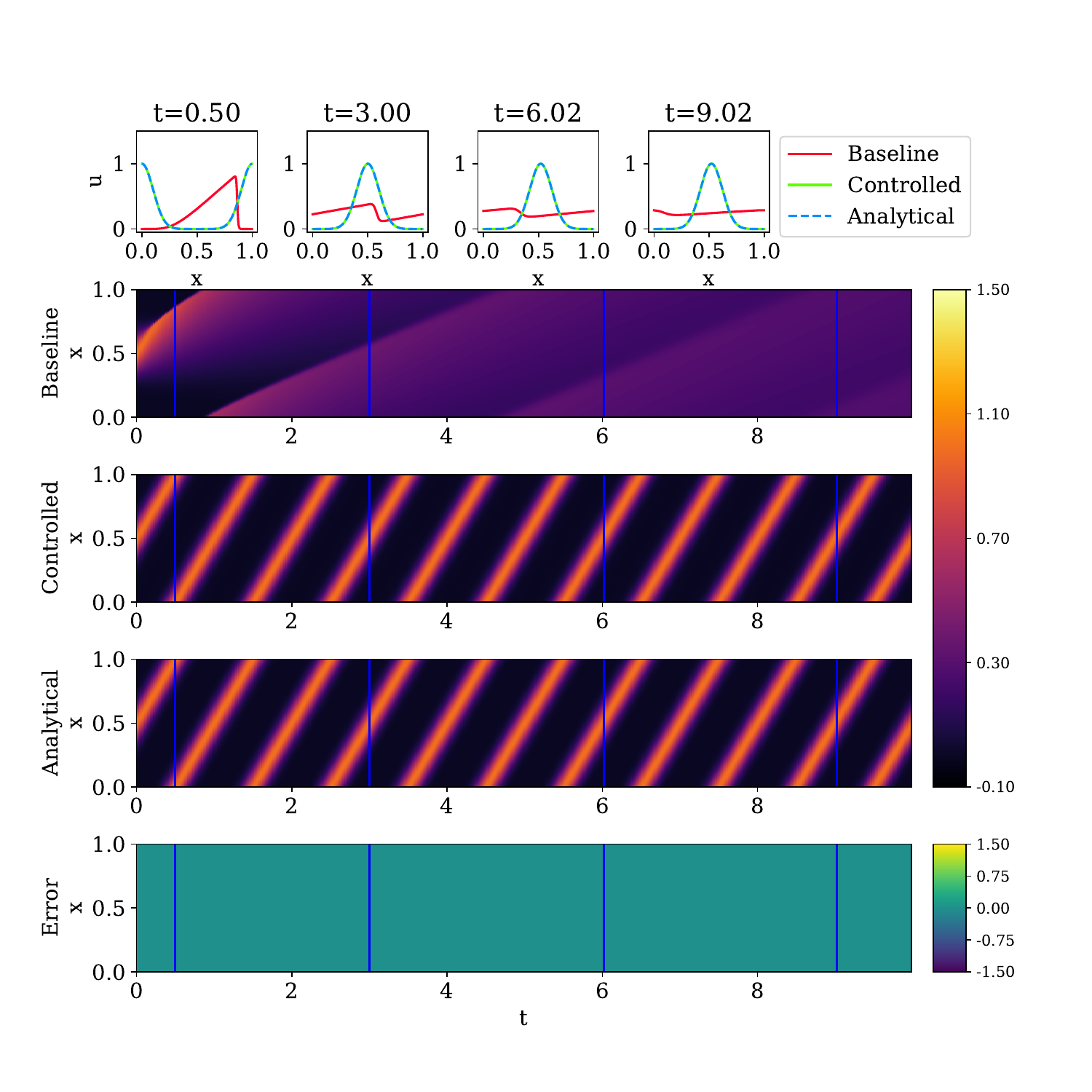}
\caption{In-sample performance:  \emph{a priori} ML (left) and DPM (right) controllers targeting an advecting Gaussian function. Top to bottom in each group: Instantaneous snapshots, uncontrolled baseline solution, controlled Burgers solution, analytical target, and error in the controlled solution.}
\label{fig:burgers_1}
\end{figure}

Models (a) and (c) are now tested over a longer window $t_\mathrm{test}/T=10$.
Figure~\ref{fig:burgers_1} displays the space-time evolution of the uncontrolled (baseline) and controlled viscous Burgers' equation for these models over this longer test duration.  The DPM-controlled solution is comparable to the target solution, even well beyond the training time window, while the quality of the \emph{a priori} control degrades for $t/T>2$.
The space-time mean-squared error (MSE) of the controlled solution,
\begin{equation}
  \epsilon = \frac{1}{Lt_\mathrm{test}}\iint (u-u^e)\,\d t\,\d x,
  \label{eq:MSE}
\end{equation}
is 4.37\,\% for the \emph{a priori} ML controller and 0.00002\,\% for the DPM controller for this in-sample test.

\begin{figure}
\centering
\includegraphics[width=0.48\linewidth]{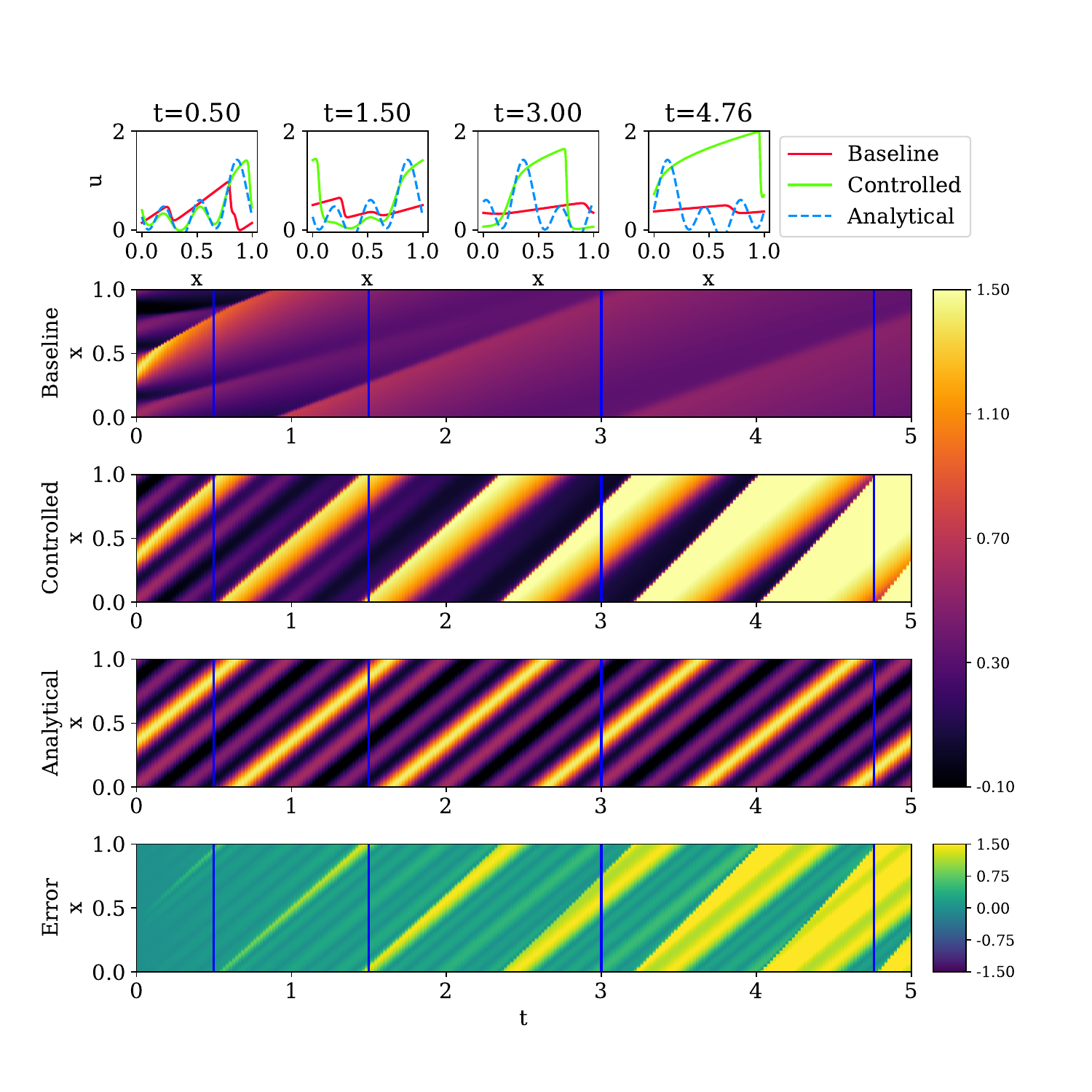}
\includegraphics[width=0.48\linewidth]{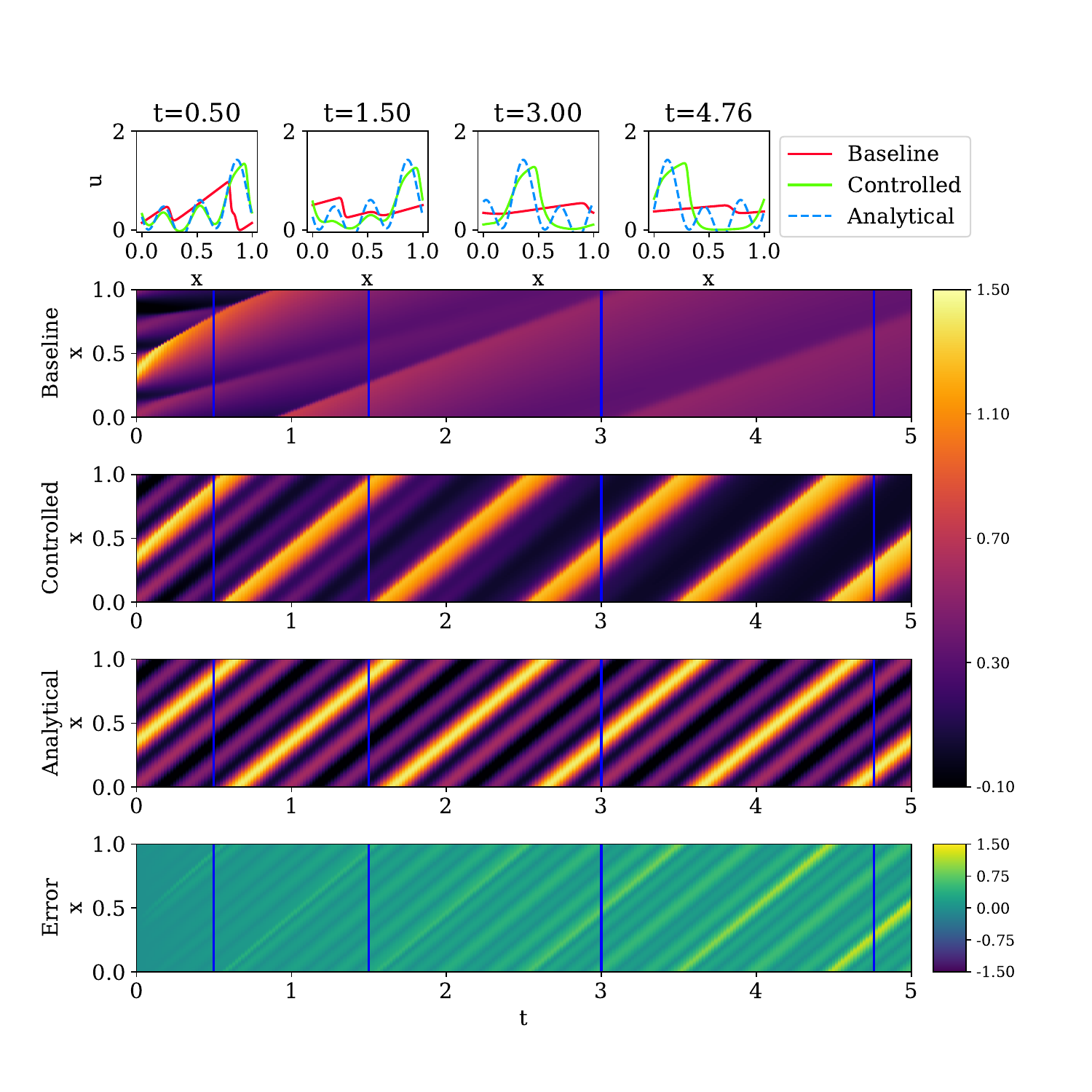}
\caption{Out-of-sample performance:  \emph{a priori} ML (left) and DPM (right) controllers targeting a multiscale Fourier function. Top to bottom in each group: Instantaneous snapshots, uncontrolled baseline solution, controlled Burgers solution, analytical target, and error in the controlled solution.}
\label{fig:burgers_2}
\end{figure}

Out-of-sample testing targets are generated using Fourier series with random coefficients  $(a_l,b_l,c)$\cite{GENEVA2020109056},
\begin{align}
  u_0(x) &= \frac{2w(x)}{\max_x|w(x)|} + c, \nonumber\\
  w(x) &= a_0 + \sum_{l=1}^{L} a_l \sin(2\pi l x) + b_l \cos(2\pi l x),\\
  a_l, b_l &\sim \mathcal{N}(0,1),\ L=4,\ \text{and}\  c\sim \mathcal{U} (0,1), \nonumber
\end{align}
where $\mathcal{N}(0,1)$ is the normal distribution with mean 0 and variance 1, and $\mathcal{U}(0,1)$ is a uniform distribution with minimum 0 and maximum 1. This is a stringent test, for the functional form of the corresponding exact source term is significantly different from that of the in-sample Gaussian target.
Figure \ref{fig:burgers_2} shows control results for the out-of-sample target. The controlled systems gradually diverge for both control methods, occurring more slowly for DPM than for \emph{a priori} ML, with mean-squared errors 58.03\,\% for \emph{a priori} ML control and 8.70\,\% for DPM control. The latter has almost 7$\times$ better stability for this out-of-sample test.

\subsubsection{Influence of the DPM training window} \label{subsubsec:tau/T}

The choice of the DPM training window $\tau$ can strongly affect training convergence and long-time stability. Despite this,  previous applications of DPM \cite{Sirignano2020DPM, MacArt2021Embedded} were limited to fixed $\tau$. We now assess its influence by training DPM models over $\tau/T \in [2\times 10^{-4}, 1]$ for advecting Gaussian targets.

\begin{figure}[t]
    \centering
    \includegraphics[width=\textwidth]{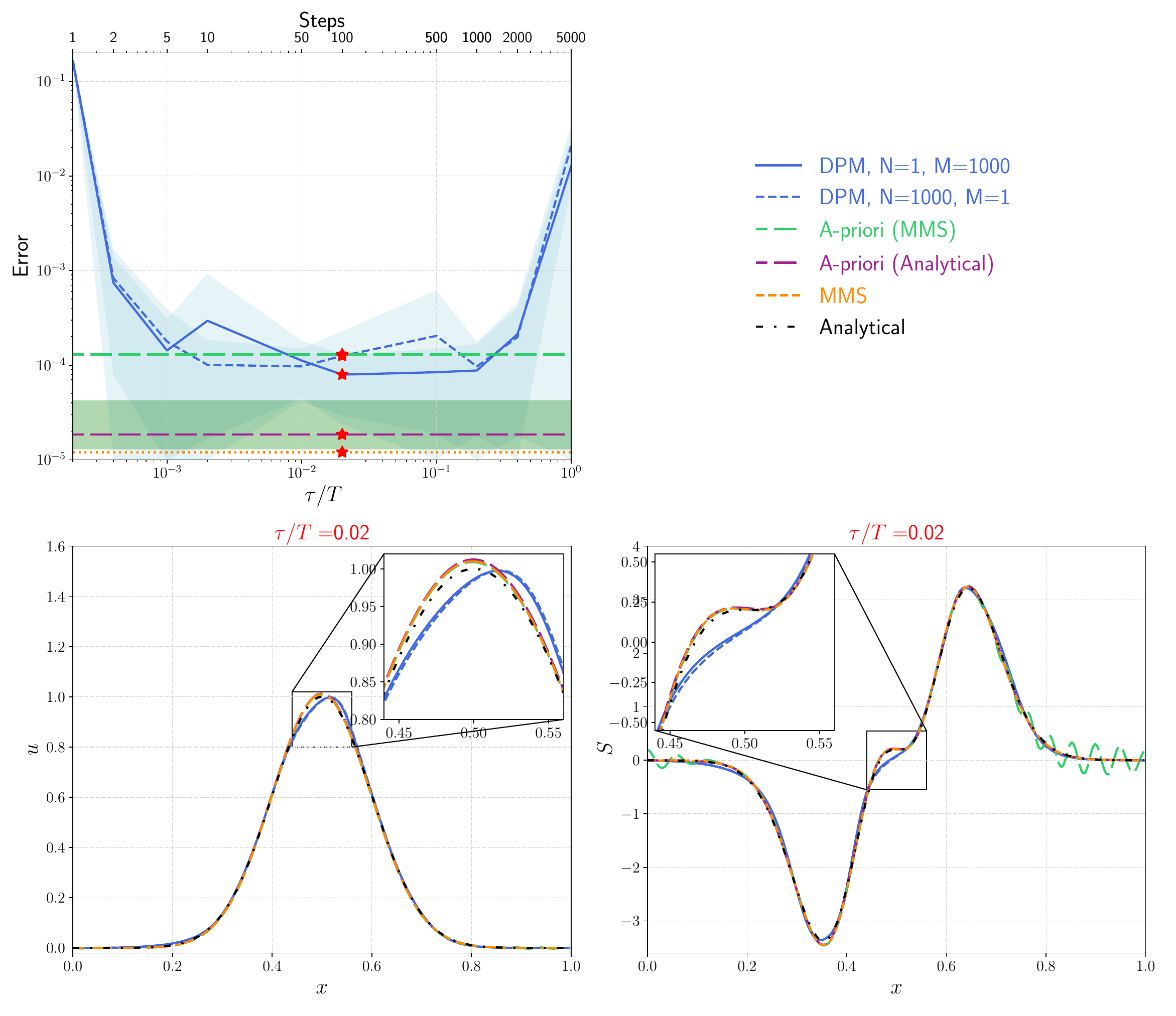}
    \caption{Top: Error of DPM, \emph{a priori}, and MMS controlled solutions, showing means and confidence intervals $\pm\sigma$ obtained using 10 random processes. Bottom: Instantaneous comparisons of analytical and controlled solutions (left) and control terms (right) at $t/T=1$; DPM models used training window $\tau/T=0.02$.
    }
    \label{fig:BurEr}
\end{figure}

Figure~\ref{fig:BurEr} displays the in-sample MSE \eqref{eq:MSE} of these solutions versus $\tau/T$; ten randomly initialized training processes were used to obtain confidence intervals $\pm\sigma$. Instantaneous snapshots of controlled solutions and forcing terms for $\tau/T=0.02$ are also shown.
The DPM control error is minimized for $\tau/T$ ranging from $10^{-3}$ ($\tau/\Delta t=5$) to $10^{-1}$ ($\tau/\Delta t=500$).
Large testing errors occur for $\tau/T < 10^{-3}$ ($\tau/\Delta t<5$), for which the observation windows are too short to fully observe the system's dynamics, and for $\tau/T\approx 1$ ($\tau/\Delta t>5{,}000$), for which the adjoint magnitudes are significantly larger than the Jacobian determinant, leading to roundoff error accumulation \cite{vishnampet2015}.

Figure~\ref{fig:BurEr} also compares the in-sample control performance of \emph{a priori}-trained neural networks targeting \eqref{eq:MMSsource} with numerically evaluated derivatives (labeled ``A-priori MMS''), \emph{a priori}-trained models targeting the analytic source term, and the numerically solved MMS solution without a neural network (i.e., evaluating \eqref{eq:MMSsource} numerically during simulation). Their errors are generally lower than the DPM error, though it is important to recall the DPM controller's better out-of-sample performance (Figure~\ref{fig:burgers_2}).
Furthermore, controllers with analytic source terms are extremely rare in practice, rendering \emph{a priori} control difficult or impossible, while DPM is capable of targeting any quantity derived from the flow solution.



For the subsequent applications to laminar flow control, the DPM training window is set to $\tau/\Delta t=10, \tau/T=0.016$ for the incompressible Navier--Stokes equations (Section~\ref{sec:DPMvsDRL}) and $\tau/\Delta t=50, \tau/T=0.005$ for the compressible equations (Section~\ref{sec:PyflowCLDPM}). These were found to provide adequate training for this viscous Burgers' equation example while maintaining reasonable computational cost.

\section{DPM versus DRL for active flow control} \label{sec:DPMvsDRL}

We compare the control performance and training cost of DPM- and DRL-based flow controllers for drag reduction in two-dimensional (2D), incompressible, laminar flows over a confined cylinder on unstructured meshes. The flow configuration and numerical solver are identical to those used for DRL control by Rabault \emph{et al.} \cite{rabault2019artificial,Rabault_2019}; the unstructured-mesh DPM implementation is new.

\subsection{Numerical simulation of confined cylinder flow} \label{subsec:FeniCS}

\begin{figure}
    \centering
    \includegraphics[width=1.0\textwidth]{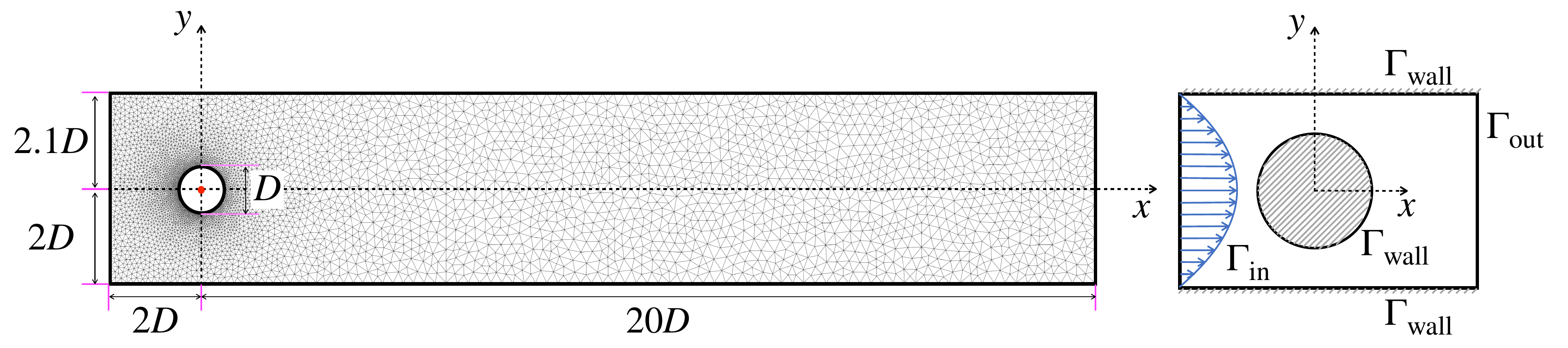}
    \caption{Left: Geometrical configuration of 2D confined flow, illustrating the unstructured mesh density. Right: Boundary conditions (not to scale).}
    \label{fig:cfcyl_geo}
\end{figure}

The flow configuration is adapted from the benchmark computations of Sch{\"a}fer et al.~\cite{schafer1996benchmark}, in which a cylinder of diameter $D$ is situated in a two-dimensional domain of size $(L,H)=(22D, 4.1D)$ that is open at its streamwise ends and bounded by no-slip walls in the cross-stream direction. The coordinate system origin $(x,y)=(0,0)$ is placed at the cylinder center, and the cylinder is shifted vertically  $0.05D$ from the  channel centerline.  This geometry and its boundaries are depicted in \figref{fig:cfcyl_geo}.

The flow is governed by the unsteady, incompressible, dimensionless Navier--Stokes equations,
 \begin{align}
    \pp{u_i}{t} +  u_j \pp{u_i}{x_j}  + \pp{p}{x_i} - \frac{1}{\Rey} \pp{^2 u_i}{x_j \partial x_j} &= f_i(p,\theta) \label{eq:momentum}\\
    \pp{u_k}{x_k} &= 0, \label{eq:mass}
\end{align}
 where $x_j$ and $t$ are the dimensionless space and time coordinates, $u_i(x_j,t)$ and $p(x_j,t)$ are the  dimensionless flow velocity and pressure, and $f_i(p;\theta)$ are control functions of the local pressure with model parameters $\theta$ (described in Section~\ref{sec:confined_control}). The dimensional length, velocity, and time scales are $D$, $\Bar{U}$, and $D/\Bar{U}$, respectively, where $\Bar{U}$ is the bulk velocity. All models are trained for Reynolds number $\Rey = \Bar{U} D/ \nu=100$ flow, where $\nu=\mu/\rho$ is the kinematic viscosity, $\mu$ is the constant dynamic viscosity, and $\rho$ is the constant density. In Section~\ref{sec:PyflowCLDPM}, the assumptions of constant viscosity and density are relaxed for compressible flows, and extrapolation of the learned models to higher Reynolds numbers is tested.

No-slip  boundary conditions  ($u=v=0$) are  imposed  at  the  top  and  bottom  walls  and along the cylinder surface. 
The inflow velocity profile at $\Gamma_{\text{in}}$ is
 \begin{equation}
    u_{\Gamma_{\mathrm{in}}}=u_{1}(x=-2D,y) = -4U_m(y-2.1D)(y+2D)/H^2,
 \end{equation}
where 
 \begin{equation}
     U_m = \frac{3}{2}\Bar{U} =\frac{3}{2} \frac{1}{H} \int_{-2D}^{2.1D} u_{\Gamma_{\mathrm{in}}}dy 
 \end{equation}
is the centerline inlet streamwise velocity (the maximum inflow velocity). The  convective outflow boundary condition on $\Gamma_{\text{out}}$ assumes zero streamwise velocity derivatives at the outlet: $\partial u_1/\partial x_1|_{\Gamma_\mathrm{out}}=0$.

The computational domain is discretized using an unstructured mesh of 9262 triangular cells with local refinement near the cylinder surface as shown in \figref{fig:cfcyl_geo}.
The minimum and maximum cell diameters (twice the circumradius) are 0.00193 and 0.0366, respectively. 
The flow equations are solved on this mesh using the finite-element method (FEM) and backward-Euler time integration using the \textit{FEniCS} framework~\cite{LoggMardalEtAl2012}. The dimensionless time step size is $\Delta t = 5\times 10^{-4}$. More details on the flow geometry and numerical methods may be found in \cite{rabault2019artificial}. 

For analysis, the instantaneous drag force $F_d$ and coefficient $C_d$ at the cylinder boundary $\Gamma_{\mathrm{cyl}}$ are
\begin{subequations}
\begin{align}
    F_d &= \int_{\Gamma_{\mathrm{cyl}}} (\tau_{1j} - p\delta_{1j} ) {n}_j\, ds \quad \text{and}\\
    C_d &= \frac{2F_d}{\rho\Bar{U}^2D},
\end{align}
\end{subequations}
where $\tau_{ij}$ is the shear-stress tensor, $\delta_{ij}$ is the Kronecker delta function, and $n_j$ is the local unit normal vector  at the cylinder surface. Time-averaged quantities $\langle\cdot\rangle$ are obtained by integrating over nodal values.

\subsection{Control framework} \label{sec:confined_control}

\begin{figure}[t]
    \centering
    \includegraphics[width=0.6\textwidth]{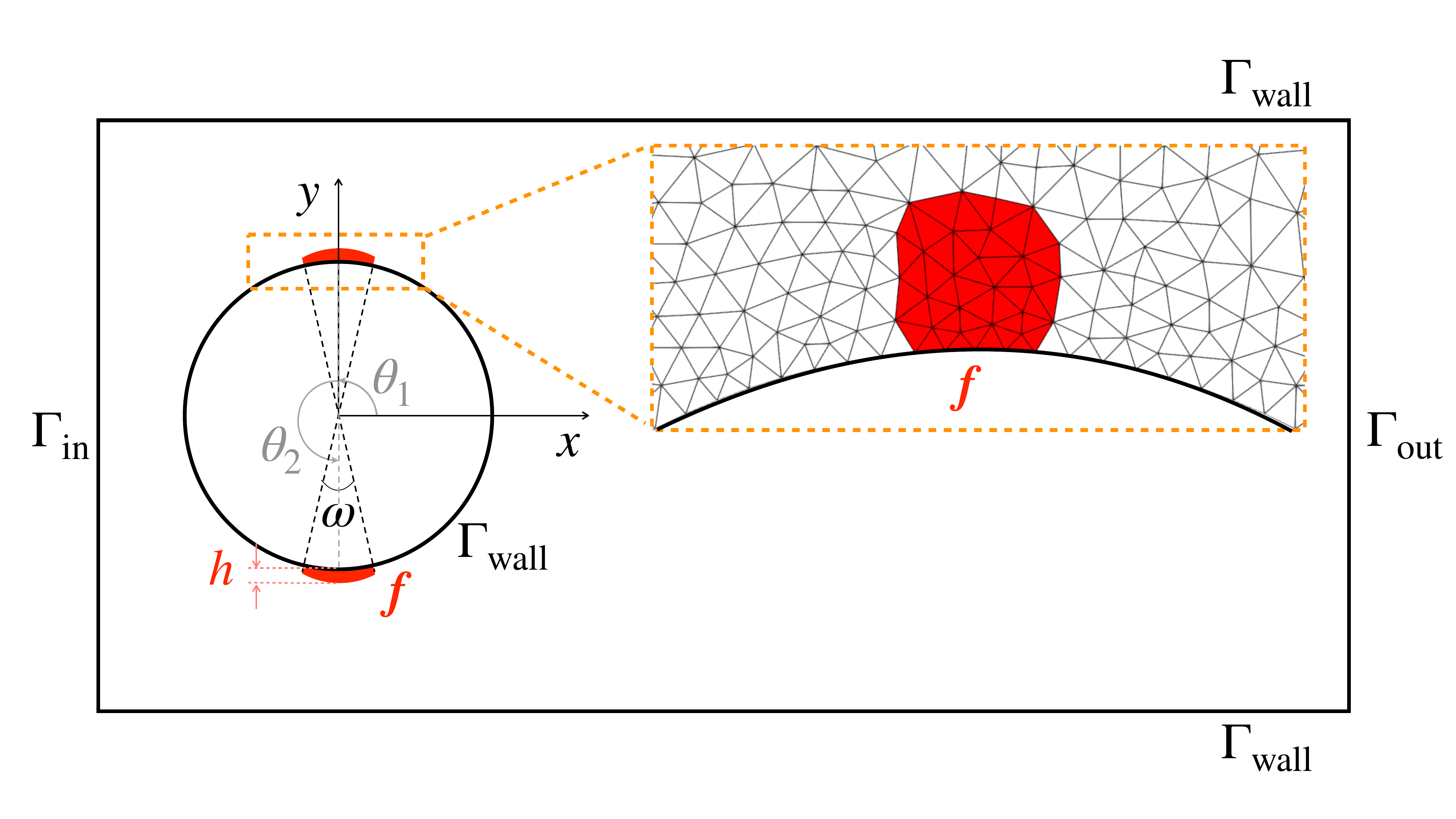}
    \vspace{-0.5cm}
    \caption{Location of the body-force actuators (red) in the 2D confined cylinder configuration. The pressure sensors for actuation are located within the same regions.}
    \label{fig:jets}
\end{figure}

The control objective is to minimize the time-averaged drag coefficient $\langle C_d \rangle$ by applying body forces $f_i$  at the cylinder surface; these are designed to emulate synthetic jets as employed in recent DRL examples~\cite{rabault2019artificial}. The body forces are centered at azimuthal angles $\theta_1=90^\circ$ and $\theta_2=270^\circ$ with width $\omega=10^\circ$ and thickness $h=0.1D$, shown in red in \figref{fig:jets}. 
On the discrete FEM grid, the body forces are applied within elements inside this control region.

The forcing terms $f_i(p;\theta)$ are approximated by feed-forward neural networks comprising either one layer with 100 hidden units (L1H100) or two layers with 512 hidden units each (L2H512). Both neural networks use hyperbolic-tangent activation functions.
The NN input is the local pressure $p$ within the control region; this is in contrast to the global velocity sensor used in recent DRL-based control \cite{rabault2019artificial}. The NN outputs (control actuations) are deterministic for DPM and stochastic for DRL. 
The two optimization methods and their respective objective functions are now discussed.

\subsubsection{DPM-based control framework} \label{sec:DPM2dunconfind}

In the PDE-constrained DL framework, the time-averaged objective $\Bar{J}$ is
\begin{equation} \label{eq:DPMobj}
    \Bar{J} (u, p, \theta) = \int_{\tau_\DPMsub} J(u,p,\theta)\, dt = \int_{\tau_\DPMsub} \Vert C_d(u) \Vert^2_{2} + \beta \Vert f(p,\theta) \Vert^2_{2}\, dt,
\end{equation}
where $\tau_\DPMsub=10\Delta t$ is the DPM training time window, and $\beta=0.1$ is a penalty coefficient to minimize large control energy expenditures. The forcing term $f$ is explicitly a function of the NN parameters $\theta$, and the velocity and pressure are implicitly functions of $\theta$ via the PDEs.
The control NN is defined in the FEM space with globally uniform weights and biases (defined on a zero-degree polynomial function space), whose values are initialized by sampling from a standard normal distribution. We consider only the L1H100 network for DPM; at each node, its input is the scalar pressure, and its output is the vector body force, resulting in 402 global NN parameters (weights and biases) to be optimized. Averaging over each control cell, this amounts to 4.6 parameters per cell that require training. More details on constructing the NN in FEM space can be found in Mitusch \emph{et al.}~\cite{Mitusch2021Data}.

The adjoint variables needed for optimization are computed using \textit{Dolfin-adjoint}~\cite{Mitusch2019}, an open-source, discrete-adjoint solver for differentiable forward-PDE models. 
The NN parameters are updated every ten forward steps using L-BFGS-B \cite{zhu1997algorithm}, a limited-memory algorithm for nonlinear optimization problems subject to simple bounds. The combination of automatic differentiation for the adjoints and L-BFGS-B for parameter updates enables efficient optimization without the need to manually implement adjoint PDEs.

\begin{figure}
\centering
\includegraphics[height=6.0cm]{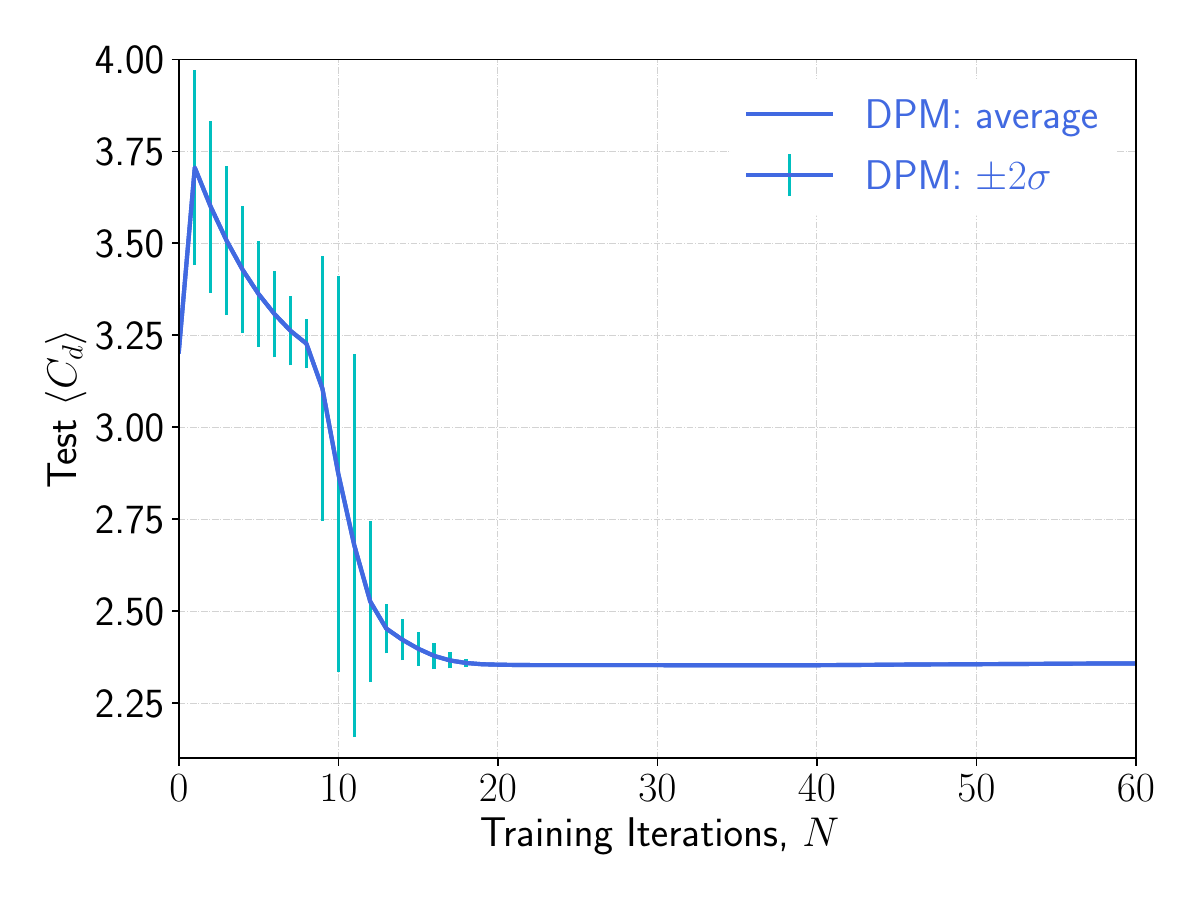}
\caption{Mean drag coefficients, obtained by averaging over $t_\mathrm{test}=9000\Delta t$, of DPM flow controllers optimized over different numbers of training iterations $N$. Test models had $<0.01\,\%$ change for $N>19$.  Confidence intervals are obtained from ten independently initialized tests per model.}
\label{fig:fenics_train}
\end{figure}

One DPM training iteration spans $\tau_\DPMsub=10\Delta t$, and iterations are not repeated ($M=1$). For the $\Rey=100$ flow considered here, the testing loss of DPM flow controllers converged after approximately $N=20$ training iterations. The convergence of the DPM-controlled mean drag coefficient, for models trained for increasing training iterations $N$, is shown in Figure~\ref{fig:fenics_train}.

\subsubsection{DRL-based control framework}

We implement DRL control following \cite{rabault2019artificial} with (a) mass-source actuators replaced by momentum actuators and (b) local pressure probes at two limited regions on the cylinder boundary instead of quasi-global velocity probes around the cylinder. 
As introduced in Section~\ref{subsec:DRL}, the DRL control framework comprises an environment with which the controller interacts---the \textit{FEniCS} simulation---and an agent trained using the PPO algorithm. 
The PPO agent is defined using  \textit{Tensorforce} \cite{tensorforce2017}, an open-source DRL platform based on \textit{TensorFlow} \cite{abadi2016tensorflow}.
The observed state is the pressure $s = (p)$, and the action is the body force $a=(f_1, f_2)$ within controlled region (red cells in \figref{fig:jets}). 
The instantaneous reward function $ r_{\hat\tau} = - \langle C_d \rangle_{\hat\tau}  - 0.2|\langle C_l \rangle_{\hat\tau}|$ combines penalties for the time-averaged lift and drag coefficient magnitudes over a sliding time window $\hat\tau=50\Delta t$.

DRL models require ``actor'' and ``critic'' networks. For the actor network, the input pressure is defined on zero-degree polynomials (a constant weight) in each controlled cell, and the output body forces are defined on two-dimensional, second-degree polynomials within each controlled cell, which are uniquely specified by 12 function weights per cell. Over the 87 controlled cells, this results in 114,244 total parameters for L1H100-based actor networks and 843,284 total parameters for L2H512-based actor networks. A given model's critic network uses the same structure as its actor network;
therefore, the total number of parameters is $\sim$228k for L1H100-based models and $\sim$1.7m for L2H512-based models.
If each controlled cell is regarded as a control point, then L1H100-based models have approximately 2.6k parameters per control point, and L2H512-based models have approximately 19.4k parameters per control point, both of which are significantly larger than the number of parameters per control point for the DPM model.


One DRL training iteration spans $\tau_\DRLsub=4000\Delta t$, corresponding to approximately  6.5  vortex  shedding  periods, after which the actor and critic NNs are updated over 25 optimization sub-iterations. A smoothly changing action function \cite{Rabault_2019} is implemented to avoid flow property jumps. DRL model convergence requires approximately $N=200$ training iterations, after which drag reduction ceases.


\subsection{DPM versus DRL control performance}

The influence of model complexity on the efficacy of the learned DPM and DRL controllers is assessed in Section~\ref{sec:complexity}. Having globally uniform parameters, the DPM models are $O(10^3)$--$O(10^4)$ times simpler than the locally defined DRL models.  The influence of the training extent on convergence and training cost is analyzed in Section~\ref{sec:cost}.

\subsubsection{Control efficacy versus model complexity} \label{sec:complexity}

\begin{figure}
    \centering
    \includegraphics[width=0.85\textwidth, ]{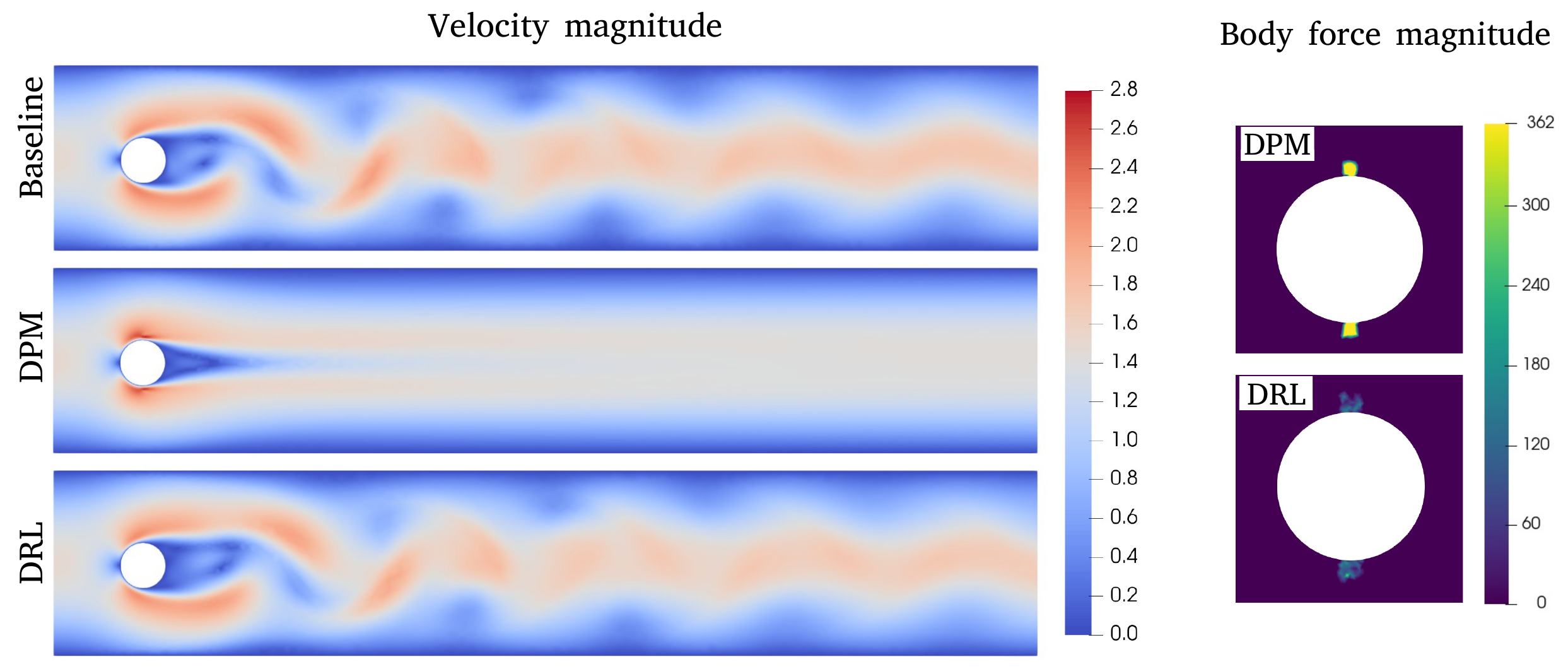}
    \caption{Left: Instantaneous velocity-magnitude snapshots for the baseline uncontrolled $\Rey=100$ flow, DPM-controlled flow (L1H100; 4.6 parameters per point), and DRL-controlled flow (L2H512; 19.4k parameters per point). Right: Instantaneous body-force control magnitudes.}
    \label{fig:u_cmp}
\end{figure}

This section compares a L1H100-based DPM model to L1H100- and L2H512-based DRL models trained for $N=200$ iterations.
Figure~\ref{fig:u_cmp} shows instantaneous snapshots of velocity magnitude for the baseline $\Rey=100$ flow, DPM-controlled flow, and L2H512 DRL-controlled flow along with the corresponding instantaneous body-force magnitudes. The DPM controller eliminates the dominant vortex shedding mode, while the DRL-controlled flow is visually similar to the baseline flow. The learned body-force terms differ significantly, with the DPM body forces having larger magnitude and greater uniformity than the DRL body forces. The form of the DPM control terms for compressible flows is analyzed in Section~\ref{sec:interpretability}.

\begin{figure}
    \centering
    \includegraphics[width=1.0\linewidth]{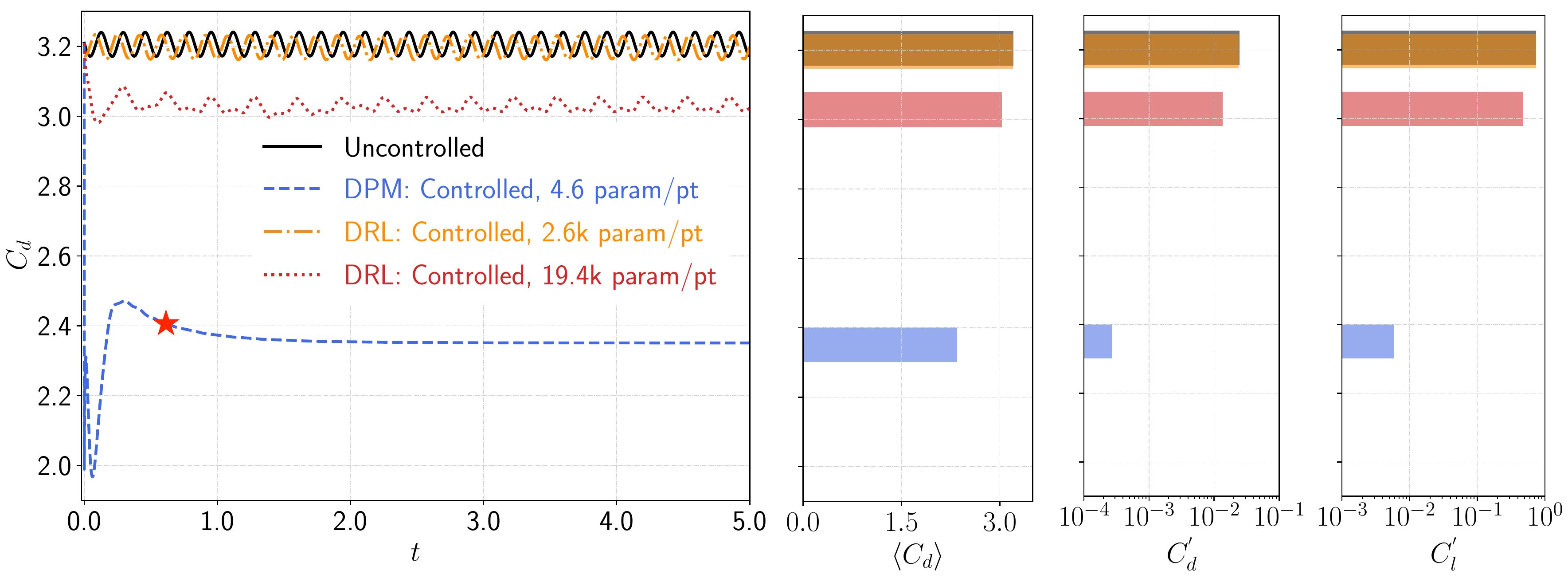}
    \caption{Left: Time evolution of uncontrolled and (in-sample) controlled instantaneous drag coefficient. Right: Mean and RMS drag coefficient and mean lift coefficient for flow times $t\geq 2.5$. The red star indicates the time after which the DPM $C_d'<5\,\%$.}
    \label{fig:complexity}
\end{figure}

\begin{table}
\centering
\caption{Control performance of DPM and DRL controllers for confined cylinder flow: Time-averaged and RMS drag coefficient, RMS lift coefficient, and Strouhal number. ``Mag.'' indicates magnitude; ``Red.~(\%)'' indicates percent reduction from the baseline.} \label{tab:DPMDRLcyl}
\begin{tabular}{ccclcclcclcc}
\toprule
      \multirow{2}{*}{}  & \multicolumn{2}{c}{$\langle C_d \rangle$} &  & \multicolumn{2}{c}{$C_d'$} &  & \multicolumn{2}{c}{$C_l'$} &  & \multicolumn{2}{c}{St} \\
        & Mag.               & Red. (\%)              &  & Mag.          & Red.\ (\%)       &  & Mag.         & Red.\ (\%)        &  & Mag.       & Red.\ (\%)      \\ \midrule 
Baseline & 3.21              & -                     &  & 0.0245        & -              &  & 0.744        & -               &  & 0.32      & -             \\
DPM (L1H100)      & 2.35              & 26.7                  &  & 0.0003        & 98.9           &  & 0.006        & 99.2            &  & 0.00      & 100.0         \\
DRL (L1H100)     & 3.20              & 0.3                   &  & 0.0239        & 3.1            &  & 0.735        & 1.1             &  & 0.32      & 0.0           \\
DRL (L2H512)     & 3.03              & 5.5                   &  & 0.0142        & 44.7           &  & 0.476        & 35.9            &  & 0.28      & 12.5          \\
\bottomrule
\end{tabular}
\end{table}

\figref{fig:complexity} presents the time-series drag coefficients, time-averaged drag coefficient, and root-mean-square $(\cdot)^{'}$ drag and lift coefficients for the uncontrolled flow, the DPM-controlled flow, and DRL-controlled flow for both networks. The efficacy of the single-layer DPM model is evident: it achieves 26.7\,\% drag reduction and 99\,\% RMS reduction from the baseline flow, while the two-layer DRL model achieves 5.5\,\% drag reduction and $\sim40$\,\% RMS reduction, and the single-layer DRL model is ineffective. Table~\ref{tab:DPMDRLcyl} summarizes the models' control performance, including the Strouhal number reduction, where $\mathrm{St} = fD/\Bar{U}$, and $f$ is the vortex-shedding frequency obtained from the Fourier transform of the instantaneous lift coefficient. The DRL models' drag reduction performance is comparable to those of Rabault \emph{et al.}\cite{rabault2019artificial, Rabault_2019}, with the present use of momentum sources rather than mass sources being the only significant difference.

That the 402-parameter DPM model achieves approximately five times the drag reduction of the 1.7m-parameter DRL model is striking, though not entirely unanticipated. The DPM training directly couples model optimization to the governing PDEs, which are in some sense the most efficient representation of the system's dynamics, though the adjoint equation's local linearization requires careful choice of the optimization window (\S\ref{subsubsec:tau/T}). This leads to high ``sample efficiency'' of the DPM training algorithm. In contrast, DRL approximates the flow dynamics using the critic network, which is potentially expensive to train and is not guaranteed to perfectly represent the controlled system. The increased training cost of DRL, discussed next, and its potentially reduced out-of-sample control efficacy are major drawbacks in configurations for which adjoint-based optimization could be performed.

\subsubsection{Control efficacy versus training cost} \label{sec:cost}

The influence of training cost on DPM and DRL controller efficacy is now assessed. The significant algorithmic differences between the two methods complicate cost comparisons; in the interest of fairness, we present results for (1) converged training loss $\langle C_d\rangle_\mathrm{train}$, obtained by averaging over the respective $\tau_\DPMsub$ or $\tau_\DRLsub$ for the final training iteration, and (2) equal training iterations ($N$).

Training cost is measured as wall time on a single AMD Ryzen Threadripper 5945WX CPU core averaged over ten independent simulations.\footnote{Both models' training can be parallelized and accelerated using GPUs, but their differing parallel efficiency further complicates cost comparisons, thus we compare only single-core performance.} DPM training required 37 seconds per iteration (3.7 seconds per $\Delta t$), and DRL training required 218 seconds per iteration (0.0545 seconds per $\Delta t$). Though the per-iteration DPM training cost is higher, largely due to the cost of constructing and solving adjoint equations at each time step, its sample efficiency enables it to converge in fewer $N$, or, for equivalent $N$, to achieve greater control efficacy. 

Training convergence (Comparison 1) was deemed to occur for $<5\,\%$ variance in $\langle C_d\rangle_\mathrm{train}$. This occurred after $N=55$ iterations for DPM models and $N=200$ iterations for DRL models. The first set of rows in Table~\ref{tab:cost} compare the wall-time cost and mean training drag for these two models. The DPM model requires approximately 47 minutes for $N=55$ iterations, resulting in $\langle C_d\rangle_\mathrm{train}=2.74$, while the DRL model requires almost 1.5 days for $N=200$ iterations,  resulting in 10\,\% higher testing loss of $\langle C_d\rangle_\mathrm{train}=3.01$. The testing drag, evaluated over $9000\Delta t$, is $\langle C_d\rangle_\mathrm{test}= 2.95$ for DRL and $\langle C_d\rangle_\mathrm{test}= 2.35$ for DPM. For DRL, $\langle C_d\rangle_\mathrm{test}\approx\langle C_d\rangle_\mathrm{train}$ due to its training time horizon ($\tau_\DRLsub$) spanning several vortex shedding cycles. Conversely, the DPM models achieve \emph{lower} controlled drag coefficients in testing due to the relatively short $\tau_\DPMsub$. This is at least indicative of the typical stability of adjoint-trained deep learning models.

\begin{table}
  \centering
  \caption{Training cost and drag reduction performance for both DPM and DRL schemes, where the time and mean drag coefficients were obtained by averaging ten independent simulations. Average wall times are reported as d:hh:mm:ss.
  } \label{tab:cost} 
  \begin{tabular}{ccc r cc}
    \toprule
    Training  & \# Parameters & Iterations $N$ & Wall Time & $\langle C_d \rangle_\mathrm{train}$ & Converged? \\ \midrule
    \multicolumn{6}{c}{\textit{Comparison 1: Converged Training Loss}} \\
    DPM & 402      & 55    & 33:48                        & 2.74                                          & Yes         \\
    DRL & 1.7m     & 200   & 1:11:52:50                   & 3.01                                             & Yes         \\
    \midrule
    \multicolumn{6}{c}{\textit{Comparison 2: Equal Training Iterations}} \\
    DPM & 402      & 14    & 8:38                         & 2.98                                           & No          \\
    DRL & 1.7m     & 14    & 46:35                        & 3.22                                            & No          \\ \bottomrule
\end{tabular}
\end{table}

To reach $\langle C_d\rangle_\mathrm{train}\approx 3.0$, the minimum achieved by DRL for $N=200$, DPM requires only $N=14$. The second set of rows in Table~\ref{tab:cost} (Comparison 2) shows the training costs for $N=14$: DPM takes approximately nine minutes; DRL takes 46 minutes and gives 8\,\% higher $\langle C_d\rangle_\mathrm{train}= 3.22$. The DRL testing drag coefficient ($\langle C_d\rangle_\mathrm{test}= 3.22$) is again similar to its training drag coefficient, while the DPM testing drag coefficient ($\langle C_d\rangle_\mathrm{test}=2.42$) is again lower than its training drag coefficient.

It is difficult to understate the efficiency of adjoint-based optimization compared to deep reinforcement learning. A fully converged, adjoint-trained DPM model requires as few as $N=20$ training iterations of duration $\tau_\DPMsub=10\Delta t$: 200 total simulation time steps. A converged DRL model---that achieves only one-fifth the drag reduction of the DPM model---requires at least $N=200$ iterations of duration $\tau_\DRLsub=4000\Delta t$. Adjoint-based optimization therefore converges approximately \emph{4000 times faster}, in terms of simulation time steps, than deep reinforcement learning and achieves superior control performance.

\section{Analysis of DPM control} \label{sec:PyflowCLDPM}
In this section, we apply DPM for drag reduction to the control of unconfined flows over cylinders governed by the compressible Navier--Stokes equations. The numerical solver and flow configuration are introduced in Sections~\ref{subsec:numerics2dconfind}--\ref{sec:DPM_control_compressible}. DPM controllers are analyzed and tested in Sections~\ref{subsec:DPM2dconfind}--\ref{sec:out-of-sample-DPM}.

\subsection{Governing equations and numerical methods } \label{subsec:numerics2dconfind}

We solve the 2D, compressible, dimensionless Navier--Stokes equations in conservative form,
\begin{align}
  \pp{ \rho}{  t} + \pp{ \rho  u_j}{  x_j} &= 0, \label{e:mass} \\
  \pp{ \rho  u_i}{  t} + \pp{ \rho  u_i  u_j}{  x_j} + \frac{1}{\Ma^2}\pp{  p}{  x_i} - \frac{1}{\Rey}\pp{ \tau_{ij}}{  x_j} &= f_{\rho u_i}, \label{e:mom} \\
  \pp{ \rho  E}{  t} + \pp{ \rho  u_j  E}{  x_j} + \frac{1}{\Ma^2}\pp{  p u_j}{  x_j} - \frac{1}{\Rey}\pp{  u_i \tau_{ij}}{  x_j} + \frac{1}{\Ma^2\Rey\Pr}\pp{  q_j}{  x_j} &= u_i f_{\rho u_i}, \label{e:energy}
\end{align}
where $\rho$ is the mass density, $E=e + u_iu_i/2$ is the total energy, $e = T/(\gamma \Ma^2)$ is the internal energy, $T$ is the temperature, $\gamma=1.4$ for calorically perfect air, and dimensionless variables are obtained by normalizing dimensional quantities $\tilde{(\cdot)}$ by a reference length $L$, freestream velocity $u_\infty$, density $\rho_\infty$, pressure $p_\infty$ and temperature $T_\infty$:
\begin{align*}
  t&=\frac{\tilde t u_\infty}{L}, &
  x_i&=\frac{\tilde x_i}{L}, &
  \rho &= \frac{\tilde\rho}{\rho_\infty}, &
  u_i &= \frac{\tilde u_i}{u_\infty}, &
  p &= \frac{\tilde p}{\gamma p_\infty}, &
  T &= \frac{\tilde T}{(\gamma-1)T_\infty}, &
  e &= \frac{\tilde e}{u_\infty^2}.
\end{align*}
This yields scaling Mach, Reynolds, and Prandtl numbers
\begin{align*}
  \Ma &= \frac{u_\infty}{\sqrt{\gamma p_\infty/\rho_\infty}}, &
  \Rey &= \frac{\rho_\infty u_\infty L}{\mu_\infty}, & &\mathrm{and} &
  \Pr &= \frac{c_p\mu_\infty}{\lambda_\infty},
\end{align*}
where $\mu_\infty$ is a dimensional reference viscosity, and $c_p$ is the specific heat at constant pressure.  The viscous-stress tensor and the heat-flux vector are
\begin{align} \label{eq:mu}
   \tau_{ij} &=  \mu\left[\left(\pp{  u_i}{  x_j} + \pp{  u_j}{  x_i}\right) - \frac{2}{3}\pp{  u_k}{  x_k}\delta_{ij}\right] +  \mu_B\pp{  u_k}{  x_k}\delta_{ij}, \\
    q_j &= - \mu\pp{  T}{  x_j},
\end{align}
with $ \mu =1.0$. The bulk viscosity $\mu_B$ is neglected for the $\Ma=0.1$ flows considered here.
The system is closed by the dimensionless ideal gas law, $  p = (\gamma-1) \rho  T/\gamma$. The momentum actuators $f_{\rho u_i}$ are defined in Section~\ref{sec:DPM_control_compressible}.

The governing equations are solved on generalized curvilinear coordinates using the transform  $(x,y)\mapsto (\xi,\eta)$. Derivatives in the computational plane $(\xi,\eta)$ are calculated  using standard fourth-order central-difference schemes and lower-order, one-sided schemes at domain boundaries.
Second derivatives are obtained by repeated application of first derivatives. Time is advanced using the fourth-order Runge--Kutta method.

Sixth-order implicit spatial filters~\cite{Lele1992} are applied at every time step to remove spurious oscillations arising from the use of central differences. A class of $2N+1$-point, maximally tridiagonal filtering schemes may be obtained from
\begin{equation}
  \zeta_f \bar\phi_{i-1} + \bar\phi_i + \zeta_f\bar\phi_{i+1} = \sum_{n=0}^N \frac{a_n}{2}\left(\phi_{i+n} + \phi_{i-n}\right),
  \label{e:filter}
\end{equation}
where $\bar\phi$ is the filtered quantity. The filter stencil reduces to lower-order, one-sided stencils near domain boundaries. Values of $a_n$ for filters of different orders may be found in \cite{Lele1992}; we use $\zeta_f=0.495$. 

\subsection{Simulation of unconfined cylinder flow} \label{subsubsec:uncflow}

The physical-space grid for an unconfined cylindrical domain is defined in terms of the computational-plane coordinates,
\begin{subequations}
\begin{align}
  x(\xi,\eta) &= R(\eta)\cos(\xi) \\
  y(\xi,\eta) &= R(\eta)\sin(\xi),
\end{align}    
\end{subequations}
with a uniform computational mesh
\begin{align*}
  \xi_i &= i\times d\xi, & d\xi &= 2\pi/N_\xi, & i&=0,\dotsc,N_\xi-1, \\
  \eta_j &= j\times d\eta, & d\eta &=1/(N_\eta-1), & j&=0,\dotsc,N_\eta.
\end{align*}

\begin{figure}[t]
  \centering
  \hspace{-0.5cm}
  \includegraphics[height=2.8cm]{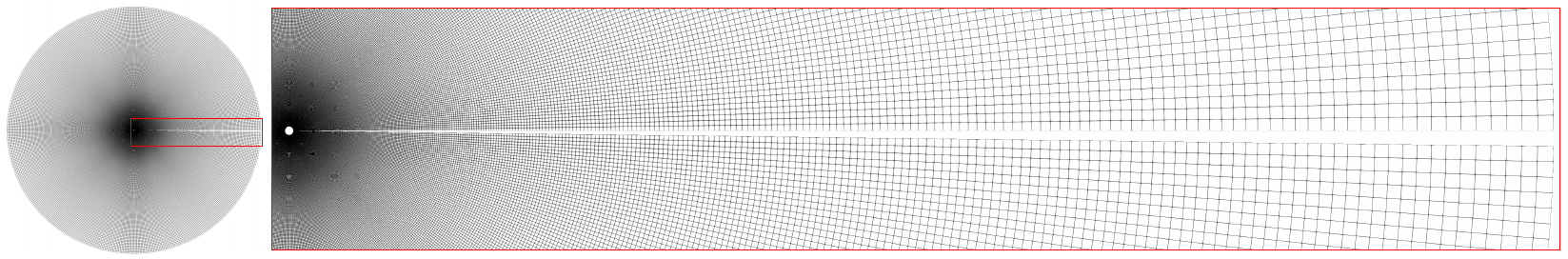}
  \caption{Left: Mesh nodes for the full domain. Right: Zoomed view of the downstream domain sector.}
  \label{fig:curvilinear}
\end{figure}

The radial grid is nonuniform using a hyperbolic-tangent stretching,
\begin{equation}
  R(\eta) = h\frac{\tanh(s_R(\eta-1))}{\tanh(s_R)} + R_\mathrm{min} + h,
\end{equation}
where $h=R_\mathrm{max}-R_\mathrm{min}$, $R_\mathrm{min}$ is the cylinder radius, $R_\mathrm{max}$ is the grid boundary radius, and $s_R$ is the stretching parameter.
The grid  is shown in \figref{fig:curvilinear} for $N_\xi=512$, $N_\eta=512$, $R_\mathrm{min}=0.5$, $R_\mathrm{max}=150$, and $s_R=6.5$. The minimum mesh spacings in the radial and azimuthal directions are $d\eta=0.0029$ and $R_\mathrm{min}d\xi=0.0061$. The azimuthal direction has a periodic boundary (the narrow gap in Figure~\ref{fig:curvilinear}).

No-slip boundary conditions are imposed on the cylinder wall. Far-field boundaries are imposed using absorbing layers; these add source terms $\sigma (\bQ_\mathrm{ref}-\bQ)$ to the governing equations, with conserved quantities $\bQ=(\rho, \rho u_i, \rho E)$ and a boundary proximity function
\begin{equation}
  \sigma(x,y) = \left\{ \begin{array}{c l}
    \alpha\left(1 - \frac{d((x,y),\partial^e\Omega)}{\delta_\alpha}\right)^p & \mathrm{for}\ \ d((x,y),\partial^e\Omega)<\delta_\alpha, \\
    0 & \mathrm{otherwise}.
  \end{array} \right.
  \label{eq:farfield}
\end{equation}
In \eqref{eq:farfield}, $d((x,y),\delta^e\Omega)$ is the distance from a point $(x,y)$ to the boundary $\delta^e\Omega$, $\delta_\alpha=1$ is the absorbing layer thickness, $\alpha=10$ is the source-term strength, and $p=3$ is its polynomial order.
Dirichlet conditions $\bQ=\bQ_\mathrm{ref}$ are then applied on $\partial^e\Omega$.

\begin{figure}[t]
    \centering
    \includegraphics[width=0.95\linewidth]{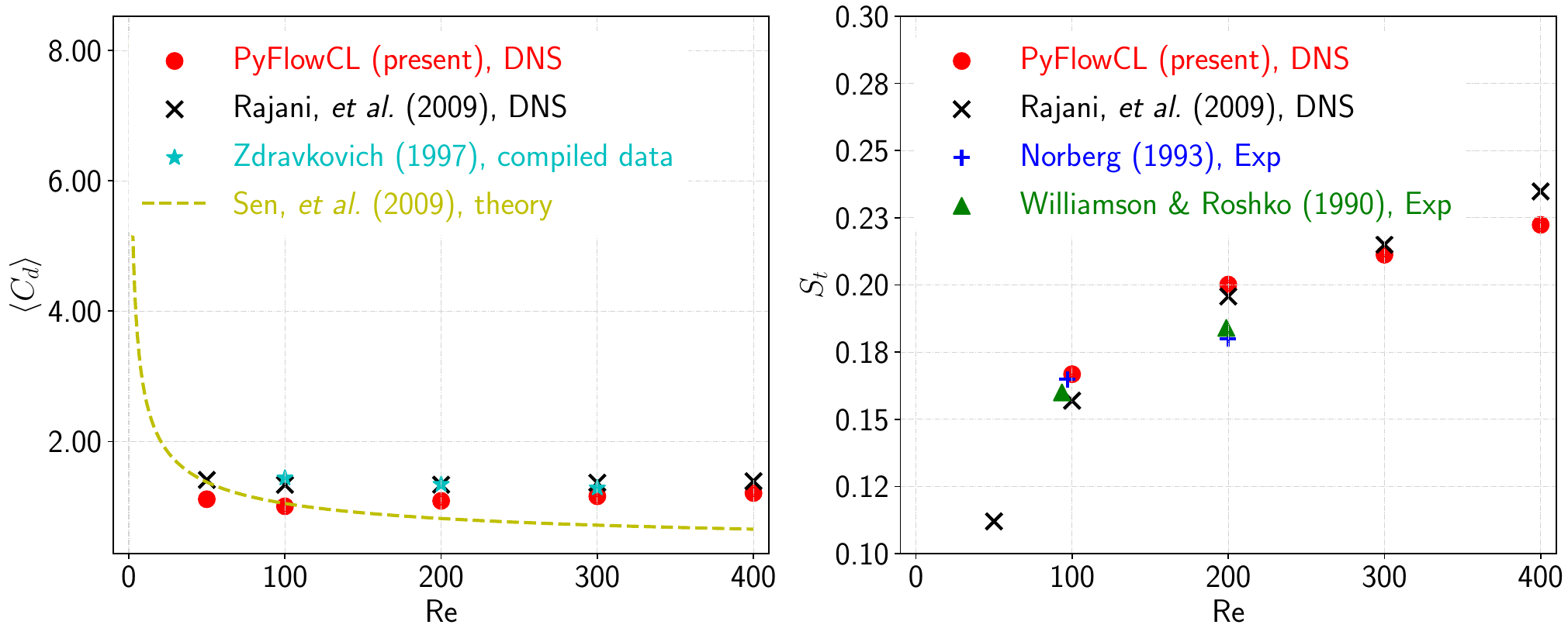} 
\caption{Mean drag coefficient $\langle C_d \rangle$ (left) and Strouhal number $S_t$ (right) at different Reynolds numbers. Data in Zdravkovich (1997)\cite{zdravkovich1997flow} is compiled from experimental, DNS and analytical theories.}\label{fig:mCdSt}
\end{figure}

The dimensionless time step size is $\Delta t = 3\times 10^{-4}$, which corresponds to CFL numbers between 0.35 and 0.45. The flow is first simulated for $t\in[0,300]$, after which time-averaged statistics (e.g., mean drag coefficient $\langle C_d \rangle$ and Strouhal number $\mathrm{St}$) are computed for $t\in [300,360]$, representing approximately 19 vortex-shedding cycles.
Figure~\ref{fig:mCdSt} displays the overall satisfactory agreement of the present computations with published data \cite{zdravkovich1997flow,rajani2009numerical, Sen2009,norberg1993pressure,williamson1990measurements} for $\langle C_d \rangle$ and $\mathrm{St}$  for Reynolds numbers  $\Rey=[50, 100, 200, 300, 400]$.


\subsection{DPM controller} \label{sec:DPM_control_compressible}

The objective function used here is \eqref{eq:DPMobj} with actuator-penalty coefficient  $\beta=10^{-5}$. Different from Section~\ref{sec:DPMvsDRL}, the body forces are now applied along the downwind cylinder boundary rather than the upper and lower cylinder surfaces.

\begin{figure}[t]
    \centering
  \qquad
  $\vcenter{\hbox{\includegraphics[width=0.5\linewidth]{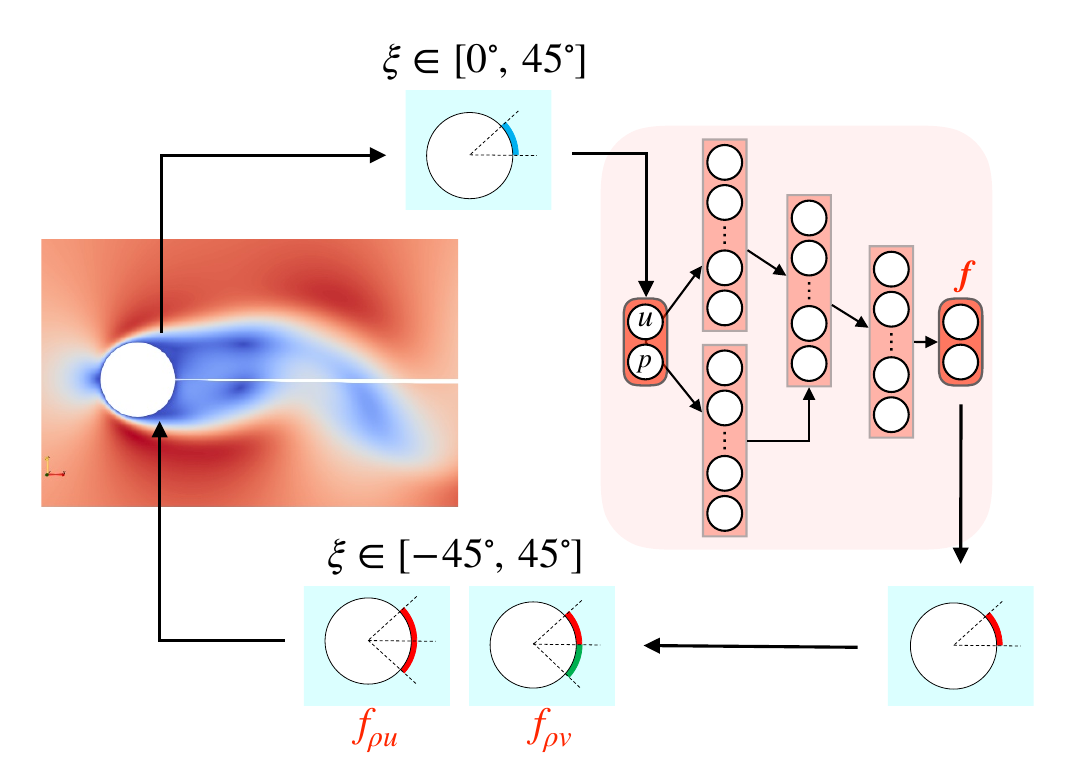} }}$ 
\caption{Illustration of the DPM-based active-control framework. }\label{fig:DPMc}
\end{figure}

The source terms $\bbf(u,p;\theta)=[f_{\rho u}, f_{\rho v}]$ are active within the region $\xi \in [-45^\circ, 45^\circ],\ \eta=0$. We model $\bbf$ using a neural network with parameters $\theta$ and inputs comprising the local velocity and pressure $(u,p)(\xi\in[0^\circ,45^\circ], \eta=0)$. The neural-network outputs are $\bbf(\xi\in[0^\circ,45^\circ], \eta=0)$, which are mirrored symmetrically ($f_{\rho u}$) and anti-symmetrically ($f_{\rho v}$) about $\xi=0^\circ$. The  fully connected neural network has one input  layer, two  consecutive  hidden  layers, a gate layer that performs an element-wise multiplication on the first hidden layer,   and one linear output layer. The hidden layers have 200 units each and use rectified linear unit (ReLU) activation functions. This control framework and neural-network structure are illustrated in \figref{fig:DPMc}.

The controller is trained for $\Rey = 100$ flow, starting from a statistically stationary initial condition $\bQ_0=(\rho_0,\rho u_0, \rho v_0, \rho E_0)$.
The neural-network parameters are updated over optimization windows $\tau=50\Delta t$; one training epoch spans $N=1000$ optimization windows without repeats ($M=1$). The influence of the number of training epochs is assessed in Section~\ref{subsec:DPM2dconfind}.
Parameter gradients are updated using the \emph{RMSprop} optimizer with constant learning rate $\alpha=10^{-4}$.
Adjoints needed for optimization are computed using automatic differentiation over the discretized forward equations using the \textit{PyTorch} package \cite{NEURIPS2019_9015}. One training epoch required approximately 8.4 wall-time hours on one 64-core AMD EPYC 7532 server node.

\subsection{Control Effectiveness} \label{subsec:DPM2dconfind}

\begin{figure}
 \centering
\includegraphics[width=0.9\linewidth]{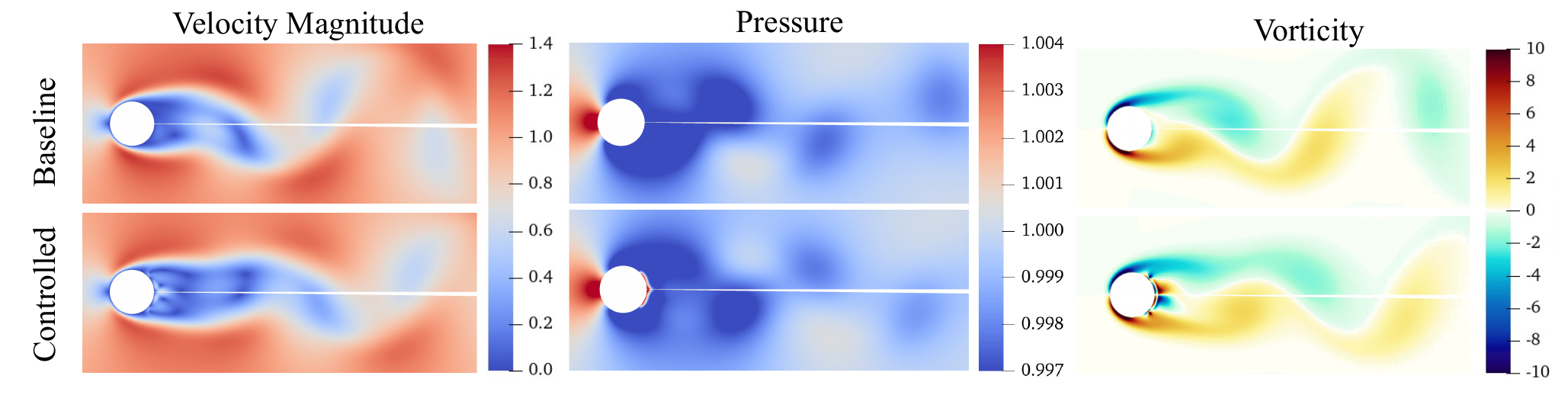}
\caption{Velocity magnitude, pressure, and vorticity magnitude snapshots for the baseline and online-DPM controlled flow. The controlled fields for the simplified-DPM controller are visually similar to those for the online-DPM control.}\label{fig:VelVorRe1}
\end{figure}

The DPM controller is trained for two epochs and tested for $t_\mathrm{test}=60$ time units.
\figref{fig:VelVorRe1} illustrates the qualitative changes to the velocity magnitude, pressure, and vorticity in the actively controlled flow. New vortical structures originate in the controlled region, representing counter-rotating vorticity to the dominant vortex-shedding modes, and the vortical shedding in the wake is visually suppressed.
The actuation forces cause higher pressures within the controlled region; these partially offset the upstream stagnation pressure and reduce the overall pressure drag  (the main drag source for this flow). 

\begin{figure}
\centering
\includegraphics[height=5.35cm]{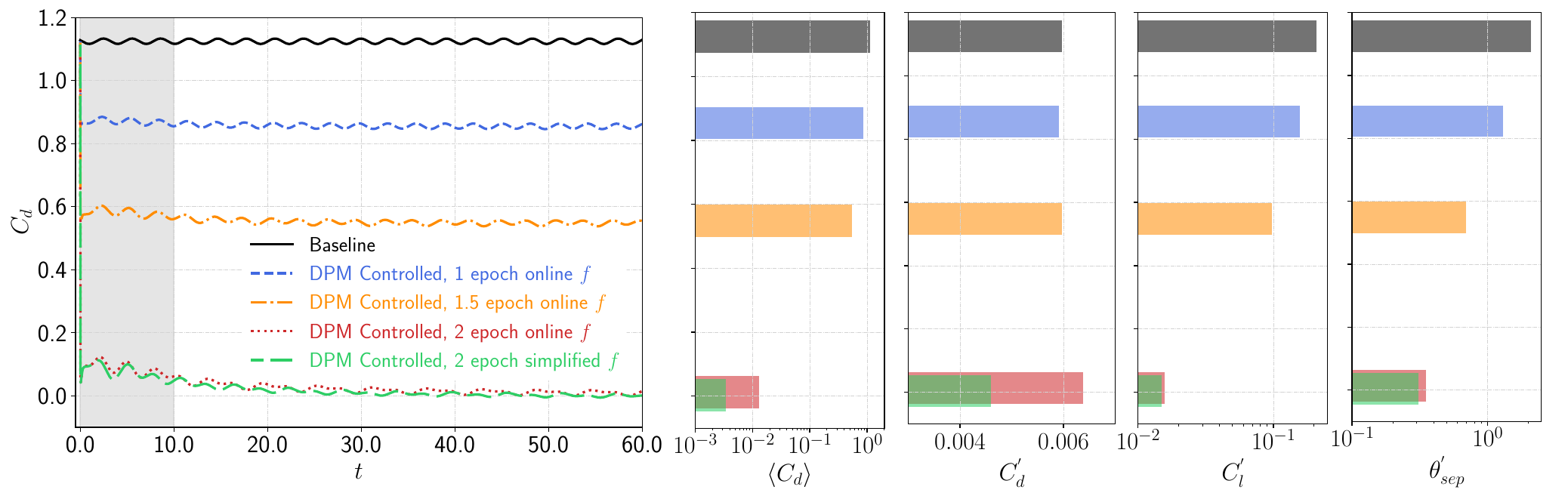}
 \caption{Time-series drag coefficient $C_d$ and statistical measures of flows controlled by online and simplified DPM actuators.} \label{fig:InterRe1}
\end{figure}

The time-averaged drag $\langle C_d \rangle$ and root-mean-square drag $C_d^{'}$, lift coefficient $C_l^{'}$, and separation angle $\theta^{'}_{\mathrm{sep}}$ are presented in \figref{fig:InterRe1} and Table~\ref{tab:DPMuncyl} for DPM-trained models for one, one and a half, and two training epochs. Also shown is the performance of simplified, constant actuator $\bbf_\mathrm{simp} = \langle \bbf(u,p;\theta) \rangle_{0.05t_{\mathrm{test}}}$ obtained by averaging the ``online'' actuator $\bbf(u,p;\theta)$ over $0.05t_{\mathrm{test}}$. Significantly, the simplified actuator does not have any inputs. The variance in the online actuator is low (Section~\ref{sec:interpretability}), hence this constant actuator is surprisingly effective.

The DPM-trained controllers significantly reduce the instantaneous drag, with a 95\,\% reduction occurring within $t=0.5$ for the two-epoch-trained model and reaching 99\,\% reduction around $t=20$. The long-time control performance improves between one and two training epochs but does not change significantly with further training. The models trained here achieve overall superior drag-reduction than the DPM model in Section~\ref{sec:DPMvsDRL} due to the wider control region and deeper neural network.

\begin{table}
\caption{Control performance of online and simplified (constant) DPM controllers for unconfined cylinder flow: Time-averaged and RMS drag coefficient, RMS lift coefficient, and separation angle RMS. ``Mag.'' indicates magnitude; ``Red.~(\%)'' indicates percent reduction from the baseline.} \label{tab:DPMuncyl}
\centering
\begin{tabular}{cccccccccccccc}
\toprule
        \multirow{2}{*}{ }           & \multirow{2}{*}{\# Ep.} & \multicolumn{2}{c}{$\langle C_d \rangle$} & \multicolumn{2}{c}{$C_d'$} & \multicolumn{2}{c}{$C_l'$} & \multicolumn{2}{c}{$\theta'_{\mathrm{sep}}$} \\ 
                     &                        & Mag.                & Red.\ (\%)              & Mag.           & Red.\ (\%)      & Mag.          & Red.\ (\%)       & Mag.                  & Red.\ (\%)                 \\ \midrule 
Uncontrolled              & --                      & 1.12               & -                    & 0.00596       & -             & 0.2096       & -              & 2.09                 & -                        \\
\multirow{3}{*}{Online DPM} & 1                      & 0.85               & 23.9                 & 0.00591       & 6.3           & 0.1571       & 24.6           & 1.31                 & 37.4                     \\
                     & 1.5                    & 0.55               & 51.2                 & 0.00596       & 21.7          & 0.0977       & 53.1           & 0.70                 & 66.7                     \\
                     & 2                      & 0.13               & 98.8                 & 0.00638       & 57.9          & 0.0156       & 92.1           & 0.35                 & 83.2                     \\
Simplified           & --                     & 0.003              & 99.7                 & 0.00460       & 20.6          & 0.0150       & 92.5           & 0.31                 & 85.1                     \\ \bottomrule
\end{tabular}
\end{table}


There are also evident effects of the controllers on flow oscillations.
The lift fluctuations $C_l'$ are reduced with increasing training epochs, with approximately $O(10)$ reduction after two training epochs. Likewise, the RMS fluctuations in the flow-separation angle $\theta_{sep}^{'}$ decrease with training. While these properties were not directly targeted for optimization, they could be included in loss functions if desired, for example, for resonance and flutter control in fluid-structure interaction problems.



\subsection{Interpretability} \label{sec:interpretability}

\begin{figure}[t]
\centering
\includegraphics[width=\textwidth]{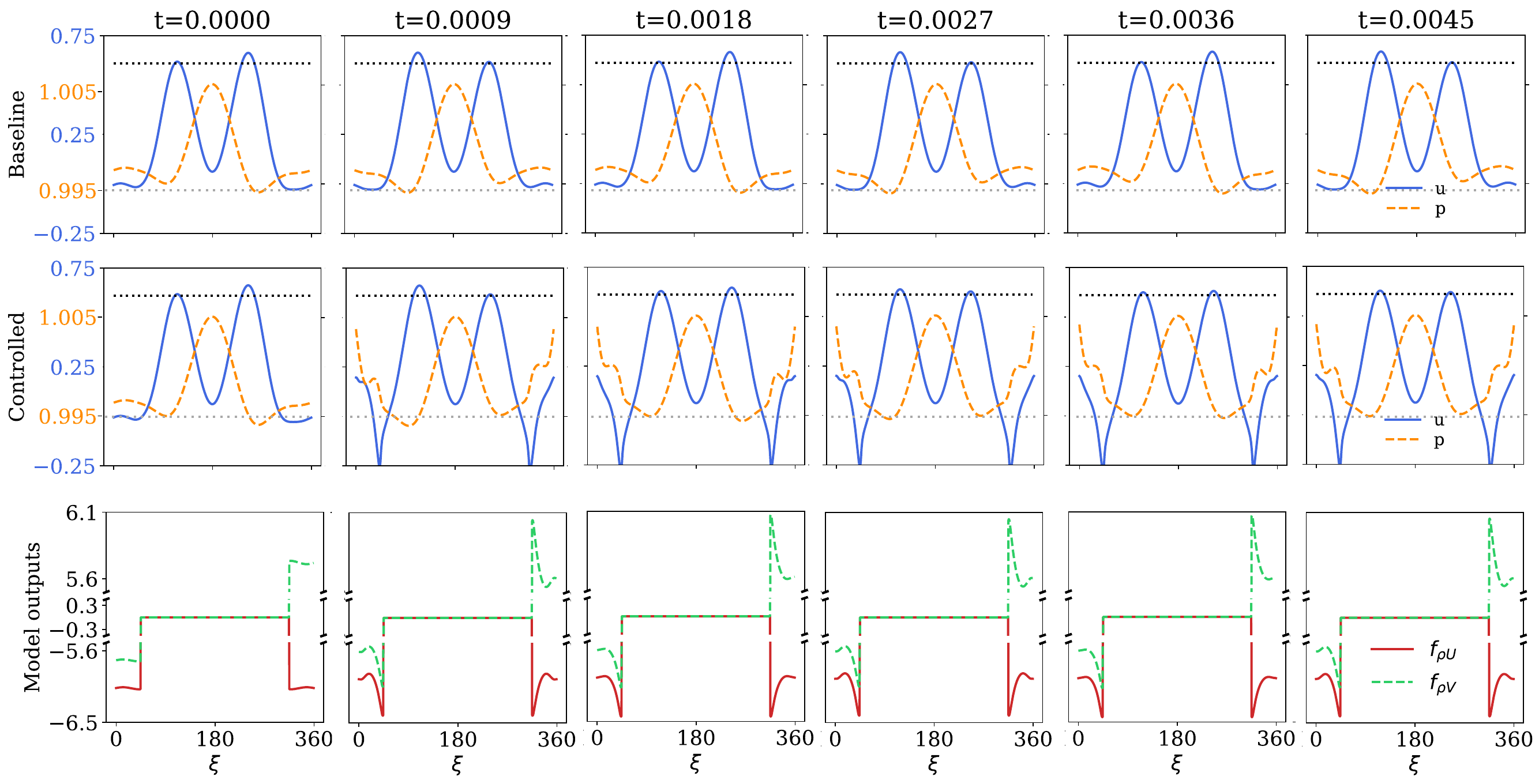}
\vspace{-6mm}
 \caption{Sequential snapshots of velocity (blue), pressure (orange) and control forces (red and green) along the cylinder boundary. The control-force axes are cut to show the full extents.} \label{fig:upfRe1}
\end{figure}

Figure~\ref{fig:upfRe1} shows snapshots of the velocity and pressure at the cylinder surface in the uncontrolled and controlled flow (using the two-epoch-trained model) at instants spaced by $3\Delta t$ after the onset of control actuation. The baseline flow has clear oscillations of the maximum velocity $u_{\mathrm{max}}$ and minimum pressure $p_{\mathrm{min}}$ that alternate between peaks at $\xi=100^\circ$ and $\xi=260^\circ$ (slightly upstream of the vertical points).
The two-epoch-trained model suppresses the flow oscillations within $t=6\Delta t$, leading to a quasi-stable flow pattern in which $u_{\mathrm{max}}\approx0.61$ and $p_{\mathrm{min}}\approx0.0995$. Within the control region ($\xi\in[-45^\circ,45^\circ]$), the velocity is negative and the pressure is positive, which together cause the observed drag reduction.
The actuator forces, also shown in \figref{fig:upfRe1}, exhibit only slight variations after $t=6\Delta t$; this motivates the simplified (constant) actuator assessed in Section~\ref{subsec:DPM2dconfind}.

\subsection{Out-of-sample performance}\label{sec:out-of-sample-DPM}

\begin{figure}
  \centering
  \vspace{-2mm}
\includegraphics[width=0.475\textwidth]{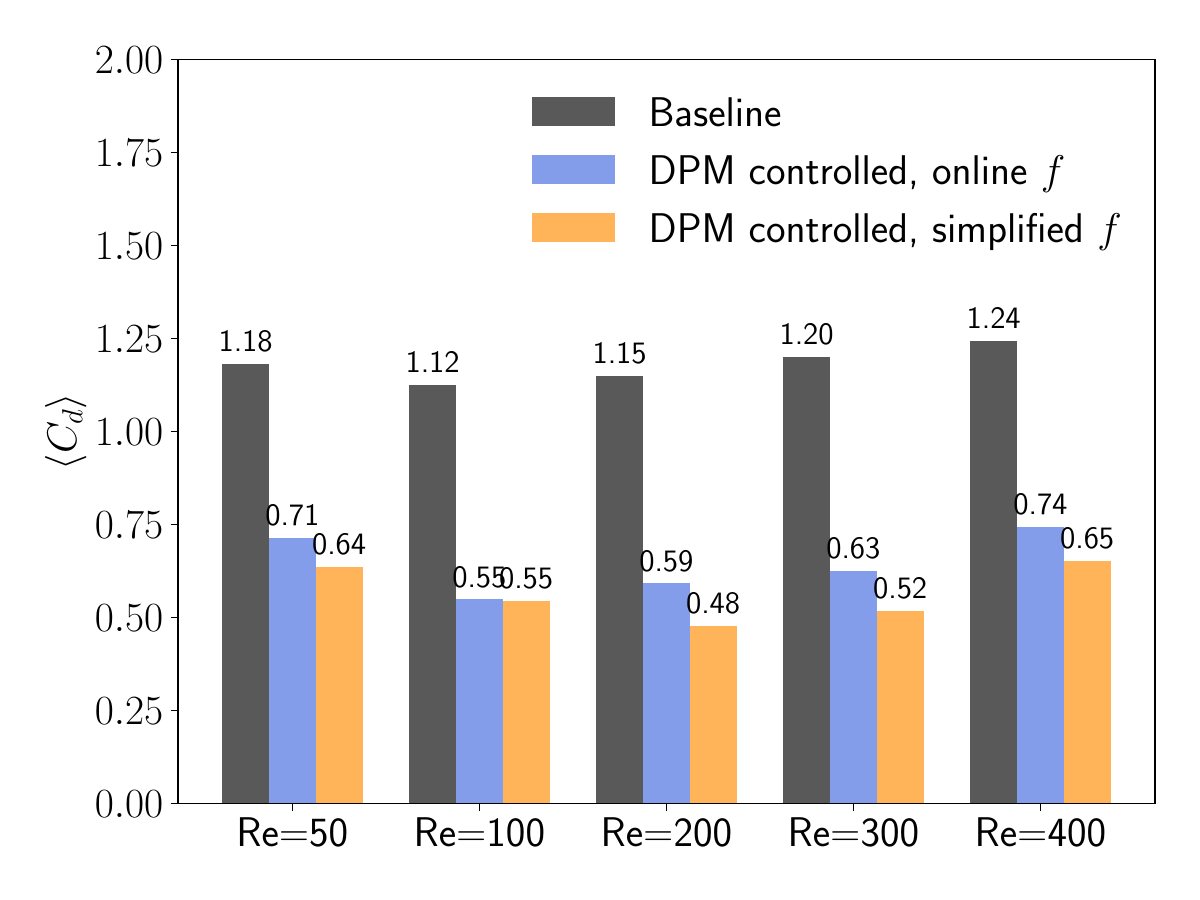}
\includegraphics[width=0.475\textwidth]{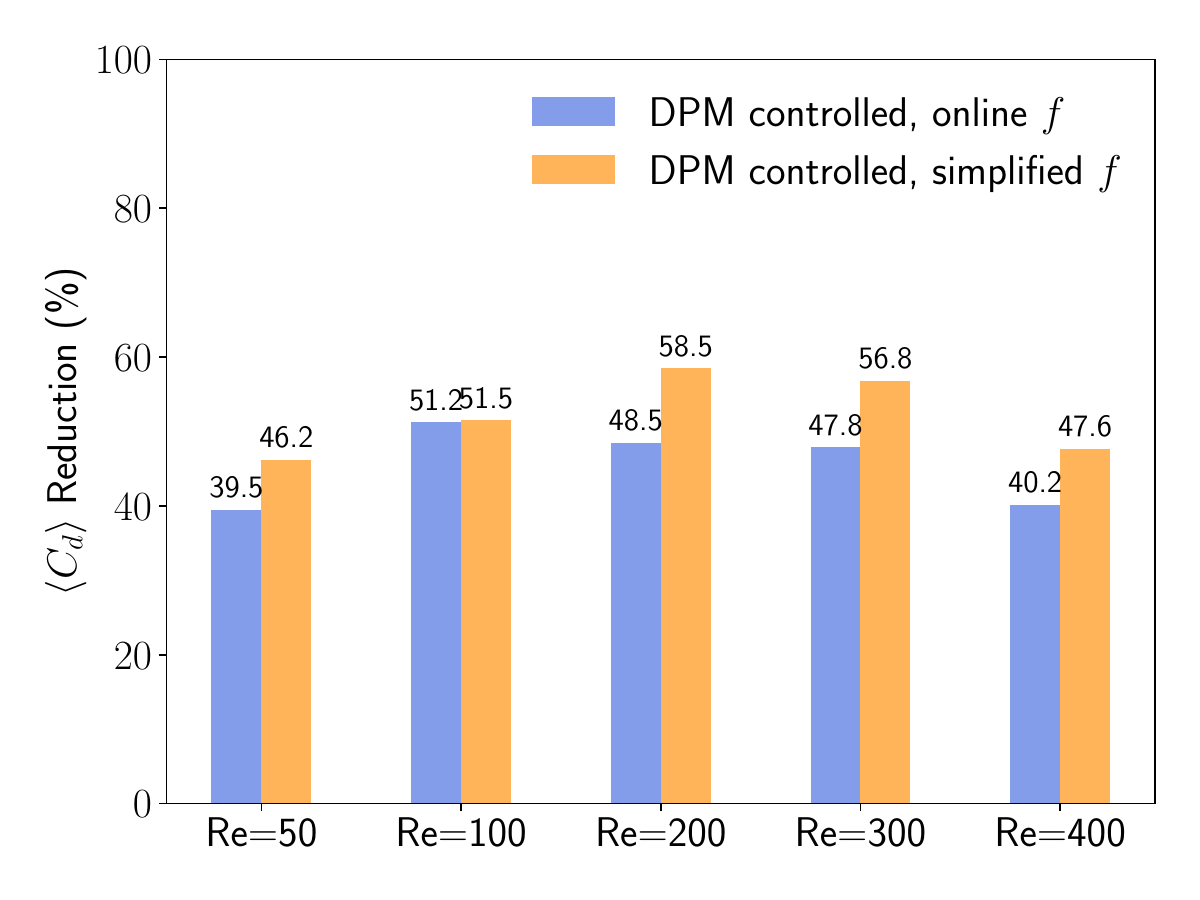}
  \vspace{-2mm}
 \caption{Drag reduction using the online ($\Rey=100$-trained) and  constant (simplified) DPM  controllers for out-of-sample Reynolds numbers.} \label{fig:CmpRe1234}
\end{figure}

The effectiveness of the one-epoch, $\Rey=100$ trained DPM controller and its constant, simplified variant are now assessed for out-of-sample Reynolds numbers between 50 to 400.
Figure~\ref{fig:CmpRe1234} shows the mean drag and reduction percentage for these cases.
The online and simplified models both have reasonably good control performance over this range of Reynolds numbers, with the simplified model performing approximately 8\,\% better on average (though it was not as effective at minimizing RMS quantities).
The online model has out-of-sample drag reduction between 39.5\,\% ($\Rey=50$) and 48.5\,\% ($\Rey=200$), though its extrapolative capacity diminishes as the training-to-testing Reynolds number difference increases. This is particularly evident for $\Rey=50$, which is less than the critical Reynolds number for vortex shedding.

Most importantly, the DPM-trained controller is stable and relatively accurate when applied to out-of-sample Reynolds numbers. In general, the effectiveness of an active flow controller can be expected to degrade when applied far from its development regime. Here, the DPM-trained controller is successful because its training closely couples with the underlying flow physics, at least as represented by the Navier--Stokes PDEs.

\section{Conclusion}\label{sec:Conclusion}

We apply adjoint-based, PDE-constrained deep learning to develop closed-loop active controllers for flows. We  assess the method's efficacy, training requirements, and out-of-sample performance, compared to benchmark \emph{a priori} learning and exact control terms, using the method of manufactured solutions applied to the 1D viscous Burgers' equation. We then apply the method to drag reduction in a 2D incompressible Navier--Stokes flow over a confined cylinder, making efficacy and cost comparisons to deep reinforcement learning-trained control terms. Finally, we apply the method to 2D compressible Navier--Stokes flow over an unconfined cylinder and test the method's out-of-sample extrapolation to lower and higher Reynolds numbers.

The MMS viscous Burgers' example attempts to fully remove nonlinearity in the solution via control terms. These are modeled using neural networks, trained using either \emph{a priori} learning or adjoint-based training, and are compared to the analytic MMS control terms. For equal training loss, adjoint-based training requires 2.15\,\% the iterations of \emph{a priori} training; for equal training iterations, the adjoint-based method achieves almost two orders of magnitude lower training error. The adjoint-trained controllers produce stable and accurate long-time solutions even for challenging out-of-sample control targets. 
The adjoint-training time window significantly influences controller effectiveness; an optimum is found between five and 500 time steps for this example system.

Deep reinforcement learning has recently shown promise for flow control; we compare it to the adjoint-training approach for drag reduction around a confined cylinder at $\Rey=100$. The adjoint-trained controller reduces the mean drag approximately five times as effectively as deep reinforcement learning with a neural-network controller that is several orders of magnitude less costly to train and evaluate. This difference is unsurprising, for the adjoint-based optimization maximally leverages the PDE constraints during training.

We finally apply the adjoint-based method to compressible, unconfined flow over a cylinder at $\Ma=0.1$. An adjoint-trained model targeting drag reduction at $\Rey=100$ achieves 98.8\,\% drag reduction within two training epochs, and a simplified, constant forcing obtained from the mean outputs of the ``online'' controller is likewise effective. The $\Rey=100$-trained model is stable and successful for drag reduction between $\Rey=50$ and $\Rey=400$, with its effectiveness diminishing by less than 10\,\%  over this range.

The effectiveness of adjoint-based deep learning for laminar flow control is encouraging. Future work will focus on the model features and training methods required for  deep learning control of increasingly chaotic and nonlinear flows including turbulent, high-speed, and reacting flows. These represent significantly more challenging scenarios than the laminar flows considered here. The use of generic actuators is also limiting for applications; the possibility also exists to design realistic control actuators using PDE-constrained methods.

\section*{Acknowledgments}
The authors thank Prof.~Justin Sirignano and Dr.~Tom Hickling for helpful discussions. This work was supported by the National Science Foundation under award CBET-2215472. The authors acknowledge computational time on resources supported by the University of Notre Dame Center for Research Computing (CRC).


\bibliographystyle{my-elsarticle-num}

\begin{thebibliography}{10}
\expandafter\ifx\csname url\endcsname\relax
  \def\url#1{\texttt{#1}}\fi
\expandafter\ifx\csname urlprefix\endcsname\relax\def\urlprefix{URL }\fi
\expandafter\ifx\csname href\endcsname\relax
  \def\href#1#2{#2} \def\path#1{#1}\fi

\bibitem{ashill2001flow}
M.~Gad-el Hak, \emph{{Flow Control: Passive, Active, and Reactive Flow
  Management}}, Cambridge University Press, 2000.

\bibitem{sudin2014review}
M.~N. Sudin, M.~A. Abdullah, S.~A. Shamsuddin, F.~R. Ramli, M.~M. Tahir, Review
  of research on vehicles aerodynamic drag reduction methods,
  \emph{International Journal of Mechanical and Mechatronics Engineering}
  14~(2) (2014) 37--47.

\bibitem{wang2019investigation}
H.~Wang, W.~Gan, D.~Li, An investigation of the aerodynamic performance for a
  fuel saving double channel wing configuration, \emph{Energies} 12~(20) (2019)
  3911.

\bibitem{CATTAFESTA2008479}
L.~N. Cattafesta, Q.~Song, D.~R. Williams, C.~W. Rowley, F.~S. Alvi,
  \href{https://www.sciencedirect.com/science/article/pii/S0376042108000584}{Active
  control of flow-induced cavity oscillations}, \emph{Progress in Aerospace
  Sciences} 44~(7) (2008) 479--502.
\newblock

\bibitem{gerontakos2006dynamic}
P.~Gerontakos, T.~Lee, Dynamic stall flow control via a trailing-edge flap,
  \emph{AIAA Journal} 44~(3) (2006) 469--480.

\bibitem{bliamis2022numerical}
C.~Bliamis, Z.~Vlahostergios, D.~Misirlis, K.~Yakinthos, {Numerical Evaluation
  of Riblet Drag Reduction on a MALE UAV}, \emph{Aerospace} 9~(4) (2022) 218.

\bibitem{acarer2020peak}
S.~Acarer, {Peak lift-to-drag ratio enhancement of the DU12W262 airfoil by
  passive flow control and its impact on horizontal and vertical axis wind
  turbines}, \emph{Energy} 201 (2020) 117659.

\bibitem{SRINATH20101994}
D.~Srinath, S.~Mittal,
  \href{https://www.sciencedirect.com/science/article/pii/S002199910900638X}{An
  adjoint method for shape optimization in unsteady viscous flows},
  \emph{Journal of Computational Physics} 229~(6) (2010) 1994--2008.
\newblock

\bibitem{prandtl1904flussigkeitsbewegung}
L.~Prandtl, {{\"U}ber Flussigkeitsbewegung bei sehr kleiner Reibung}, in:
  Verhandl. III Internat. Math. Kongr. Heidelberg (1904), Leipzig, 1905, pp.
  484--491.

\bibitem{rabault2019artificial}
J.~Rabault, M.~Kuchta, A.~Jensen, U.~R{\'e}glade, N.~Cerardi, Artificial neural
  networks trained through deep reinforcement learning discover control
  strategies for active flow control, \emph{Journal of Fluid Mechanics} 865
  (2019) 281--302.

\bibitem{tang2020robust}
H.~Tang, J.~Rabault, A.~Kuhnle, Y.~Wang, T.~Wang, {Robust active flow control
  over a range of Reynolds numbers using an artificial neural network trained
  through deep reinforcement learning}, \emph{Physics of Fluids} 32~(5) (2020)
  053605.

\bibitem{roussopoulos1993feedback}
K.~Roussopoulos, Feedback control of vortex shedding at low reynolds numbers,
  \emph{Journal of Fluid Mechanics} 248 (1993) 267--296.

\bibitem{choi1993feedback}
H.~Choi, R.~Temam, P.~Moin, J.~Kim, {Feedback control for unsteady flow and its
  application to the stochastic Burgers equation}, \emph{Journal of Fluid
  Mechanics} 253 (1993) 509--543.

\bibitem{brunton2015closed}
S.~L. Brunton, B.~R. Noack, Closed-loop turbulence control: Progress and
  challenges, \emph{Applied Mechanics Reviews} 67~(5) (2015) 050801.

\bibitem{act11070201}
G.~Farrell, M.~Gibbons, T.~Persoons,
  \href{https://www.mdpi.com/2076-0825/11/7/201}{Combined passive/active flow
  control of drag and lift forces on a cylinder in crossflow using a synthetic
  jet actuator and porous coatings}, \emph{Actuators} 11~(7) (2022) 201.
\newblock

\bibitem{SCOTTCOLLIS2004237}
S.~S. Collis, R.~D. Joslin, A.~Seifert, V.~Theofilis,
  \href{https://www.sciencedirect.com/science/article/pii/S0376042104000405}{Issues
  in active flow control: theory, control, simulation, and experiment},
  \emph{Progress in Aerospace Sciences} 40~(4) (2004) 237--289.
\newblock

\bibitem{brunton2020machine}
S.~L. Brunton, B.~R. Noack, P.~Koumoutsakos, Machine learning for fluid
  mechanics, \emph{Annual Review of Fluid Mechanics} 52 (2020) 477--508.

\bibitem{portal2022cnn}
K.~Portal-Porras, U.~Fernandez-Gamiz, E.~Zulueta, A.~Ballesteros-Coll,
  A.~Zulueta, {CNN-based flow control device modelling on aerodynamic
  airfoils}, \emph{Scientific Reports} 12~(1) (2022) 8205.

\bibitem{liao2021multi}
P.~Liao, W.~Song, P.~Du, H.~Zhao, Multi-fidelity convolutional neural network
  surrogate model for aerodynamic optimization based on transfer learning,
  \emph{Physics of Fluids} 33~(12) (2021) 127121.

\bibitem{mohan2018deep}
A.~T. Mohan, D.~V. Gaitonde, {A deep learning based approach to reduced order
  modeling for turbulent flow control using LSTM neural networks}, \emph{arXiv
  preprint arXiv:1804.09269}.

\bibitem{otto2019linearly}
S.~E. Otto, C.~W. Rowley, Linearly recurrent autoencoder networks for learning
  dynamics, \emph{SIAM Journal on Applied Dynamical Systems} 18~(1) (2019)
  558--593.

\bibitem{viquerat2021direct}
J.~Viquerat, J.~Rabault, A.~Kuhnle, H.~Ghraieb, A.~Larcher, E.~Hachem, Direct
  shape optimization through deep reinforcement learning, \emph{Journal of
  Computational Physics} 428 (2021) 110080.

\bibitem{sonoda2023reinforcement}
T.~Sonoda, Z.~Liu, T.~Itoh, Y.~Hasegawa, Reinforcement learning of control
  strategies for reducing skin friction drag in a fully developed turbulent
  channel flow, \emph{Journal of Fluid Mechanics} 960 (2023) A30.

\bibitem{lee2023turbulence}
T.~Lee, J.~Kim, C.~Lee, Turbulence control for drag reduction through deep
  reinforcement learning, \emph{Physical Review Fluids} 8~(2) (2023) 024604.

\bibitem{paris2021robust}
R.~Paris, S.~Beneddine, J.~Dandois, Robust flow control and optimal sensor
  placement using deep reinforcement learning, \emph{Journal of Fluid
  Mechanics} 913 (2021) A25.

\bibitem{li2022reinforcement}
J.~Li, M.~Zhang, Reinforcement-learning-based control of confined cylinder
  wakes with stability analyses, \emph{Journal of Fluid Mechanics} 932 (2022)
  A44.

\bibitem{pino2023comparative}
F.~Pino, L.~Schena, J.~Rabault, M.~A. Mendez, Comparative analysis of machine
  learning methods for active flow control, \emph{Journal of Fluid Mechanics}
  958 (2023) A39.

\bibitem{vignon2023recent}
C.~Vignon, J.~Rabault, R.~Vinuesa, {Recent advances in applying deep
  reinforcement learning for flow control: Perspectives and future directions},
  \emph{Physics of Fluids} (2023) 031301.

\bibitem{li2017deep}
Y.~Li, Deep reinforcement learning: An overview, \emph{arXiv preprint
  arXiv:1701.07274}.

\bibitem{carnarius2013optimal}
A.~Carnarius, F.~Thiele, E.~{\"O}zkaya, A.~Nemili, N.~R. Gauger, Optimal
  control of unsteady flows using a discrete and a continuous adjoint approach,
  in: System Modeling and Optimization: 25th IFIP TC 7 Conference, CSMO 2011,
  Berlin, Germany, September 12-16, 2011, Revised Selected Papers 25, Springer,
  2013, pp. 318--327.

\bibitem{bellman1956dynamic}
R.~Bellman, Dynamic programming and lagrange multipliers, \emph{Proceedings of
  the National Academy of Sciences} 42~(10) (1956) 767--769.

\bibitem{tsiotras2017toward}
P.~Tsiotras, M.~Mesbahi, Toward an algorithmic control theory, \emph{Journal of
  Guidance, Control, and Dynamics} 40~(2) (2017) 194--196.

\bibitem{jameson1988aerodynamic}
A.~Jameson, Aerodynamic design via control theory, \emph{Journal of Scientific
  Computing} 3 (1988) 233--260.

\bibitem{reuther1999constrained}
J.~J. Reuther, A.~Jameson, J.~J. Alonso, M.~J. Rimllnger, D.~Saunders,
  Constrained multipoint aerodynamic shape optimization using an adjoint
  formulation and parallel computers, part 2, \emph{Journal of Aircraft} 36~(1)
  (1999) 61--74.

\bibitem{gorodetsky2018gradient}
A.~A. Gorodetsky, J.~D. Jakeman, Gradient-based optimization for regression in
  the functional tensor-train format, \emph{Journal of Computational Physics}
  374 (2018) 1219--1238.

\bibitem{carnarius2010adjoint}
A.~Carnarius, F.~Thiele, E.~Oezkaya, N.~R. Gauger, Adjoint approaches for
  optimal flow control, in: 5th Flow Control Conference, 2010, p. 5088.

\bibitem{kord2019Optimal}
A.~Kord, J.~Capecelatro, {Optimal perturbations for controlling the growth of a
  Rayleigh–Taylor instability}, \emph{Journal of Fluid Mechanics} 876 (2019)
  150–185.
\newblock

\bibitem{Sirignano2020DPM}
J.~Sirignano, J.~F. MacArt, J.~B. Freund,
  \href{https://www.sciencedirect.com/science/article/pii/S0021999120305854}{{DPM:
  A deep learning PDE augmentation method with application to large-eddy
  simulation}}, \emph{Journal of Computational Physics} 423 (2020) 109811.
\newblock

\bibitem{MacArt2021Embedded}
J.~F. MacArt, J.~Sirignano, J.~B. Freund, Embedded training of neural-network
  subgrid-scale turbulence models, \emph{Physical Review Fluids} 6 (2021)
  050502.
\newblock

\bibitem{sirignano2023pde}
J.~Sirignano, J.~MacArt, K.~Spiliopoulos, {PDE-constrained models with neural
  network terms: Optimization and global convergence}, \emph{Journal of
  Computational Physics} 481 (2023) 112016.

\bibitem{SuttonBarto2018rl}
R.~Sutton, A.~Barto, \emph{Reinforcement Learning: An Introduction (2nd
  Edition)}, MIT Press, Cambridge, Massachusetts, London, England, 2018.

\bibitem{konda1999actor}
V.~Konda, J.~Tsitsiklis, Actor-critic algorithms, in: {Advances in Neural
  Information Processing Systems (NIPS)}, Vol.~12, 1999.

\bibitem{schulman2017proximal}
J.~Schulman, F.~Wolski, P.~Dhariwal, A.~Radford, O.~Klimov, Proximal policy
  optimization algorithms, \emph{arXiv preprint arXiv:1707.06347}.

\bibitem{tensorforce2017}
A.~Kuhnle, M.~Schaarschmidt, K.~Fricke,
  \href{https://github.com/tensorforce/tensorforce}{{Tensorforce: A TensorFlow
  library for applied reinforcement learning}}, Web page (2017).
\newline\urlprefix\url{https://github.com/tensorforce/tensorforce}

\bibitem{NEURIPS2019_9015}
A.~Paszke, S.~Gross, F.~Massa, A.~Lerer, J.~Bradbury, G.~Chanan, T.~Killeen,
  Z.~Lin, N.~Gimelshein, L.~Antiga, A.~Desmaison, A.~Kopf, E.~Yang, Z.~DeVito,
  M.~Raison, A.~Tejani, S.~Chilamkurthy, B.~Steiner, L.~Fang, J.~Bai,
  S.~Chintala, {PyTorch: An Imperative Style, High-Performance Deep Learning
  Library}, in: H.~Wallach, H.~Larochelle, A.~Beygelzimer, F.~d\'
  Alch\'{e}-Buc, E.~Fox, R.~Garnett (Eds.), Advances in Neural Information
  Processing Systems 32, 2019, pp. 8024--8035.

\bibitem{whitham2011linear}
G.~B. Whitham, \emph{Linear and Nonlinear Waves}, John Wiley \& Sons, 2011.

\bibitem{GENEVA2020109056}
N.~Geneva, N.~Zabaras,
  \href{https://www.sciencedirect.com/science/article/pii/S0021999119307612}{Modeling
  the dynamics of \uppercase{PDE} systems with physics-constrained deep
  auto-regressive networks}, \emph{Journal of Computational Physics} 403 (2020)
  109056.
\newblock

\bibitem{vishnampet2015}
R.~Vishnampet Ganapathi~Subramanian, An exact and consistent adjoint method for
  high-fidelity discretization of the compressible flow equations, {Ph.D.
  thesis}, University of Illinois at Urbana--Champaign (2015).

\bibitem{Rabault_2019}
J.~Rabault, A.~Kuhnle, Accelerating deep reinforcement learning strategies of
  flow control through a multi-environment approach, \emph{Physics of Fluids}
  31~(9) (2019) 094105.

\bibitem{schafer1996benchmark}
M.~Sch{\"a}fer, S.~Turek, F.~Durst, E.~Krause, R.~Rannacher, \emph{Benchmark
  computations of laminar flow around a cylinder}, Springer, 1996.

\bibitem{LoggMardalEtAl2012}
A.~Logg, K.-A. Mardal, G.~N. Wells (Eds.), \emph{Automated Solution of
  Differential Equations by the Finite Element Method: The FEniCS book},
  Vol.~84, Springer Science \& Business Media, 2012.
\newblock

\bibitem{Mitusch2021Data}
S.~K. Mitusch, S.~W. Funke, M.~Kuchta, {Hybrid FEM-NN models: Combining
  artificial neural networks with the finite element method}, \emph{Journal of
  Computational Physics} 446.
\newblock

\bibitem{Mitusch2019}
S.~K. Mitusch, S.~W. Funke, J.~S. Dokken,
  \href{https://doi.org/10.21105/joss.01292}{{Dolfin-Adjoint: Automated
  Adjoints for FEniCS and Firedrake}}, \emph{Journal of Open Source Software}
  4~(38) (2019) 1292.
\newblock

\bibitem{zhu1997algorithm}
C.~Zhu, R.~H. Byrd, P.~Lu, J.~Nocedal, {Algorithm 778: L-BFGS-B: Fortran
  subroutines for large-scale bound-constrained optimization}, \emph{ACM
  Transactions on Mathematical Software} 23~(4) (1997) 550--560.

\bibitem{abadi2016tensorflow}
M.~Abadi, P.~Barham, J.~Chen, Z.~Chen, A.~Davis, J.~Dean, M.~Devin,
  S.~Ghemawat, G.~Irving, M.~Isard, et~al., Tensorflow: A system for
  large-scale machine learning, in: 12th USENIX Symposium on Operating Systems
  Design and Implementation (OSDI), 2016, pp. 265--283.

\bibitem{Lele1992}
S.~K. Lele, {Compact finite difference schemes with spectral-like resolution},
  \emph{Journal of Computational Physics} 103~(1) (1992) 16--42.

\bibitem{zdravkovich1997flow}
M.~M. Zdravkovich, \emph{Flow around circular cylinders: Applications}, Vol.~2,
  Oxford University Press, 1997.

\bibitem{rajani2009numerical}
B.~Rajani, A.~Kandasamy, S.~Majumdar, Numerical simulation of laminar flow past
  a circular cylinder, \emph{Applied Mathematical Modelling} 33~(3) (2009)
  1228--1247.

\bibitem{Sen2009}
S.~Sen, S.~Mittal, G.~Biswas, {Steady separated flow past a circular cylinder
  at low Reynolds numbers}, \emph{Journal of Fluid Mechanics} 620 (2009)
  89--119.
\newblock

\bibitem{norberg1993pressure}
C.~Norberg, Pressure forces on a circular cylinder in cross flow, in:
  Bluff-Body Wakes, Dynamics and Instabilities, Springer, 1993, pp. 275--278.

\bibitem{williamson1990measurements}
C.~Williamson, A.~Roshko, {Measurements of base pressure in the wake of a
  cylinder at low Reynolds numbers}, \emph{Zeitschrift f\"ur Flugwissenschaften
  und Weltraumforschung} 14~(1-2) (1990) 38--46.

\end{thebibliography}

\end{document}